\documentclass[11pt,a4paper]{article}
\pdfoutput=1

\usepackage{jheppub}

\usepackage{amsmath, amssymb} 
\usepackage{mathtools}
\usepackage{esint}
\usepackage{bm}
\usepackage{dsfont}
\usepackage{braket}
\usepackage{eqparbox}

\usepackage{enumitem}
\numberwithin{equation}{section}
\usepackage[dvipsnames]{xcolor}
\usepackage{tikz}
\usepackage{calc}
\usetikzlibrary{calc}
\usetikzlibrary{decorations.pathmorphing}
\usetikzlibrary{decorations.markings}

\usepackage[numbers, sort&compress]{natbib}
\usepackage{thmtools}
\newtheorem{thm}{Theorem}[section]

\usepackage{hyperref}
\hypersetup{colorlinks=true, linktoc=page, linkcolor=purple, citecolor=blue}

\newcommand{\mcF}{\mathcal{F}}

\newcommand{\mcJ}{\mathcal{J}}

\newcommand{\mcN}{\mathcal{N}}

\newcommand{\mcP}{\mathcal{P}}

\newcommand{\mcR}{\mathcal{R}}

\newcommand{\mfn}{\mathfrak{n}}

\newcommand{\bbC}{\mathbb{C}}

\newcommand{\eps}{\varepsilon}

\renewcommand{\[}{\left[}

\newcommand{\p}{\partial}

\renewcommand{\Re}{\operatorname{Re}}
\renewcommand{\Im}{\operatorname{Im}}

\newcommand{\floor}[1]{\cramped{\left\lfloor #1 \right\rfloor}}
\newcommand{\ceil}[1]{\cramped{\left\lceil #1 \right\rceil}}

\newcommand{\widesim}[2][1.5]{\mathrel{\overset{#2}{\scalebox{#1}[1]{$\sim$}}}}

\newcommand{\half}{\tfrac{1}{2}}
    
\newcommand{\no}{\nonumber}

\begin{document}

\title{Analytic bootstrap bounds on masses and spins in gravitational and non-gravitational scalar theories}

\date{\today}

\author[a]{Justin Berman}
\author[a,b]{Nicholas Geiser}

\affiliation[a]{
    Leinweber Center for Theoretical Physics, Randall Labratory of Physics\\
    University of Michigan, Ann Arbor\\
    450 Church St, Ann Arbor, MI 48109-1040, USA}
\affiliation[b]{
    Michigan Center for Applied and Interdisciplinary Mathematics\\
    University of Michigan, Ann Arbor\\
    530 Church St, Ann Arbor, MI 48109-1040, USA}

\emailAdd{jdhb@umich.edu}
\emailAdd{ngeiser@umich.edu}

\preprint{LCTP-24-25}


\abstract{We derive analytic constraints on the weakly-coupled spectrum of theories with a massless scalar under the standard assumptions of the S-matrix bootstrap program. These bootstrap bounds apply to any theory (with or without gravity) with fully crossing symmetric (i.e.\ $stu$-symmetric) four-point amplitudes and generalize results for color- or flavor-ordered (i.e.\ $su$-symmetric) planar amplitudes recently proved by one of the authors. We assume that the theory is weakly-coupled below some cut-off, that the four-point massless scalar amplitude is polynomially-bounded in the Regge limit, and that this amplitude exchanges states with a discrete set of masses and a finite set of spins at each mass level. The spins and masses must then satisfy ``Sequential Spin Constraints" (SSC) and ``Sequential Mass Constraints" (SMC). The SSC requires the lightest spin-$j$ state to be lighter than the lightest spin-$(j+1)$ state (in the $su$\nobreakdash-symmetric case) or the lightest spin-$(j+2)$ state (in the $stu$\nobreakdash-symmetric case). The SMC requires the mass of the lightest spin-$j$ state to be smaller than some non-linear function of the masses of lower-spin states. Our results also apply to super-gluon and super-graviton amplitudes stripped of their polarization dependence. In particular, the open and closed superstring spectra saturate the SSC with maximum spins ${J_{n,\text{open}} = n+1}$ and ${J_{n,\text{closed}} = 2n+2}$, respectively, at the $n^\text{th}$ mass level.}


\maketitle

\section{Introduction}
\label{sec:intro}

The S-matrix describes the scattering of particle states in relativistic quantum theories. A program of ``bootstrapping'' a unique S-matrix from fundamental physical principles such as unitarity, causality, and locality dates back to the 1950s and 1960s~\cite{Eden:1966dnq, chew1966analytic, Cappelli:2012cto}. This program long predates the modern revolution of scattering amplitudes, and in just the last few years, the S-matrix bootstrap has reemerged with many new advances~\mbox{\cite{Dixon:1996wi, Elvang:2013cua, Paulos:2016fap, Paulos:2016but, Paulos:2017fhb, Correia:2020xtr, Caron-Huot:2020cmc, Arkani-Hamed:2020blm}}. Rather than searching for a unique description of fundamental physics, the modern S\nobreakdash-matrix bootstrap program is more concerned with studying the space of consistent theories using several different techniques and a variety of simplifying assumptions. This modern literature includes studies of theories of scalars~\cite{Caron-Huot:2020cmc, Arkani-Hamed:2020blm, deRham:2017avq, Tolley:2020gtv, Bellazzini:2020cot, Chiang:2021ziz, Chiang:2022ltp, Guerrieri:2021tak,  Bellazzini:2021oaj, CarrilloGonzalez:2022fwg, Acanfora:2023axz, McPeak:2023wmq, Wan:2024eto, Berman:2024eid}, pions~\cite{Guerrieri:2020bto, He:2021eqn, Fernandez:2022kzi, Albert:2022oes, Albert:2023jtd, Albert:2023seb, Li:2023qzs, Guerrieri:2024jkn}, photons~\mbox{\cite{Henriksson:2021ymi, Alberte:2021dnj}}, fermions~\cite{Hebbar:2020ukp}, gravitons~\cite{Camanho:2014apa, Caron-Huot:2021rmr, Bern:2021ppb, deRham:2021bll, Chiang:2022jep, Haring:2022cyf, deRham:2022gfe, Caron-Huot:2024lbf, Haring:2024wyz, Hillman:2024ouy}, superstrings~\cite{Caron-Huot:2016icg, Huang:2020nqy, Guerrieri:2021ivu, Chiang:2023quf, Haring:2023zwu, Berman:2023jys, Berman:2024wyt, Albert:2024yap}, and more. For a comprehensive review of the literature through 2022, we direct the reader to the white papers~\cite{Kruczenski:2022lot, deRham:2022hpx}.

Much of this modern work is inspired by the conformal bootstrap program~\mbox{\cite{Simmons-Duffin:2015qma, Poland:2022qrs}}. One approach uses numerical semi-definite optimization to place rigorous two-sided bounds on the space of consistent effective field theories (EFTs)~\cite{Caron-Huot:2020cmc}. A complementary approach carves out the space of consistent theories, the EFT-hedron, using analytic methods~\mbox{\cite{Bellazzini:2020cot, Arkani-Hamed:2020blm}}.

Another body of modern research follows the strategy of Veneziano, Virasoro, Lovelace, Shapiro, Coon, and others from the early days of the S-matrix bootstrap program~\mbox{\cite{Veneziano:1968yb, Virasoro:1969me, Lovelace:1968kjy, Shapiro:1969km, Shapiro:1970gy, Coon:1969yw, Cappelli:2012cto}} by attempting to write down explicit analytic expressions for unitarity, causal, and local four-point amplitudes. This approach did, after all, lead to the birth of string theory~\cite{Cappelli:2012cto}. Recent results include the discovery of several seemingly-consistent deformations of four-point string amplitudes~\cite{Huang:2022mdb, Figueroa:2022onw, Geiser:2022exp, Geiser:2022icl, Cheung:2022mkw, Cheung:2023adk, Cheung:2023uwn, Cheung:2024uhn, Cheung:2024obl, Bjerrum-Bohr:2024wyw}. While some proposals have been put forward~\mbox{\cite{Maldacena:2022ckr, Wang:2024jhc}}, these deformed amplitudes do not generally arise from known underlying microscopic theories. The deformed amplitudes are typically characterized by a set of free parameters which specify a spectrum of massive states and their couplings, but for some choices of these parameters, the amplitudes exchange ghosts and violate unitarity~\mbox{\cite{Rigatos:2023asb, Rigatos:2024beq, Wang:2024wcc, Bhardwaj:2022lbz, Bhardwaj:2024klc, Mansfield:2024wjc}}. Moreover, the consistency of these deformations to higher-point amplitudes is an open question~\mbox{\cite{Arkani-Hamed:2023jwn, Geiser:2023qqq}}. Notably, each of these deformed amplitudes (including the cases when they reduce to string amplitudes) have an infinite spectrum of higher-spin massive states.

\begin{figure}
\centering
\begin{tikzpicture}
[scale=0.7]
    \coordinate (m0) at ( 0.0, 0.0);
    \coordinate (m1) at ( 1.0, 0.0);
    \coordinate (m2) at ( 2.0, 0.0);
    \coordinate (m3) at ( 3.0, 0.0);
    \coordinate (m4) at ( 4.0, 0.0);

    \coordinate (j0) at ( 0.0, 0.0);
    \coordinate (j1) at ( 0.0, 1.0);
    \coordinate (j2) at ( 0.0, 2.0);
    \coordinate (j3) at ( 0.0, 3.0);
    \coordinate (j4) at ( 0.0, 4.0);
    \coordinate (j5) at ( 0.0, 5.0);
    \coordinate (j6) at ( 0.0, 6.0);

    \node[anchor=mid] at ( 5.3, 0.0) {$\alpha' m^2$};
    \node at ( 0.0, 7.0) {$j$};
    
    \draw[->] ( 0.0, 0.0)--( 4.5, 0.0);
    \draw[->] ( 0.0, 0.0)--( 0.0, 6.5);

    \coordinate (xr) at ( 0.2, 0.0);
    \coordinate (yr) at ( 0.0, 0.2);

    \draw (m0) -- ($(m0)-(yr)$);
    \draw (m1) -- ($(m1)-(yr)$);
    \draw (m2) -- ($(m2)-(yr)$);
    \draw (m3) -- ($(m3)-(yr)$);
    \draw (m4) -- ($(m4)-(yr)$);
    
    \draw (j0) -- ($(j0)-(xr)$);
    \draw (j1) -- ($(j1)-(xr)$);
    \draw (j2) -- ($(j2)-(xr)$);
    \draw (j3) -- ($(j3)-(xr)$);
    \draw (j4) -- ($(j4)-(xr)$);
    \draw (j5) -- ($(j5)-(xr)$);
    \draw (j6) -- ($(j6)-(xr)$);
    
    \node at ($(m0)-2.5*(yr)$) {$\scriptstyle 0$};
    \node at ($(m1)-2.5*(yr)$) {$\scriptstyle 1$};
    \node at ($(m2)-2.5*(yr)$) {$\scriptstyle 2$};
    \node at ($(m3)-2.5*(yr)$) {$\scriptstyle 3$};
    \node at ($(m4)-2.5*(yr)$) {$\scriptstyle 4$};

    \node at ($(j0)-2.5*(xr)$) {$\scriptstyle 0$};
    \node at ($(j1)-2.5*(xr)$) {$\scriptstyle 1$};
    \node at ($(j2)-2.5*(xr)$) {$\scriptstyle 2$};
    \node at ($(j3)-2.5*(xr)$) {$\scriptstyle 3$};
    \node at ($(j4)-2.5*(xr)$) {$\scriptstyle 4$};
    \node at ($(j5)-2.5*(xr)$) {$\scriptstyle 5$};
    \node at ($(j6)-2.5*(xr)$) {$\scriptstyle 6$};

    \draw[color=red, thick, dashed]    ( 0.0, 1.0)--( 4.3, 5.3);
    \draw[color=orange, thick, dashed] ( 0.0, 0.0)--( 4.3, 4.3);
    \draw[color=yellow, thick, dashed] ( 1.0, 0.0)--( 4.3, 3.3);
    \draw[color=green, thick, dashed]  ( 2.0, 0.0)--( 4.3, 2.3);
    \draw[color=blue, thick, dashed]   ( 3.0, 0.0)--( 4.3, 1.3);
    \draw[color=violet, thick, dashed] ( 4.0, 0.0)--( 4.3, 0.3);

    \filldraw[color=red]    (0,1) circle (2.5pt);
    \filldraw[color=red]    (1,2) circle (2.5pt);
    \filldraw[color=red]    (2,3) circle (2.5pt);
    \filldraw[color=red]    (3,4) circle (2.5pt);
    \filldraw[color=red]    (4,5) circle (2.5pt);

    \filldraw[color=orange] (0,0) circle (2.5pt);
    \filldraw[color=orange] (1,1) circle (2.5pt);
    \filldraw[color=orange] (2,2) circle (2.5pt);
    \filldraw[color=orange] (3,3) circle (2.5pt);
    \filldraw[color=orange] (4,4) circle (2.5pt);
    
    \filldraw[color=yellow] (1,0) circle (2.5pt);
    \filldraw[color=yellow] (2,1) circle (2.5pt);
    \filldraw[color=yellow] (3,2) circle (2.5pt);
    \filldraw[color=yellow] (4,3) circle (2.5pt);
    
    \filldraw[color=green]  (2,0) circle (2.5pt);
    \filldraw[color=green]  (3,1) circle (2.5pt);
    \filldraw[color=green]  (4,2) circle (2.5pt);

    \filldraw[color=blue]   (3,0) circle (2.5pt);
    \filldraw[color=blue]   (4,1) circle (2.5pt);
    
    \filldraw[color=violet] (4,0) circle (2.5pt);
    
\end{tikzpicture}
\qquad \qquad
\begin{tikzpicture}
[scale=0.7]
    \coordinate (m0) at ( 0.0, 0.0);
    \coordinate (m1) at ( 1.0, 0.0);
    \coordinate (m2) at ( 2.0, 0.0);
    \coordinate (m3) at ( 3.0, 0.0);
    \coordinate (m4) at ( 4.0, 0.0);

    \coordinate (j0) at ( 0.0, 0.0);
    \coordinate (j1) at ( 0.0, 1.0);
    \coordinate (j2) at ( 0.0, 2.0);
    \coordinate (j3) at ( 0.0, 3.0);
    \coordinate (j4) at ( 0.0, 4.0);
    \coordinate (j5) at ( 0.0, 5.0);
    \coordinate (j6) at ( 0.0, 6.0);

    \node[anchor=mid] at ( 5.5, 0.0) {$\tfrac{1}{4} \alpha' m^2$};
    \node at ( 0.0, 7.0) {$j$};
    
    \draw[->] ( 0.0, 0.0)--( 4.5, 0.0);
    \draw[->] ( 0.0, 0.0)--( 0.0, 6.5);

    \coordinate (xr) at ( 0.2, 0.0);
    \coordinate (yr) at ( 0.0, 0.2);

    \draw (m0) -- ($(m0)-(yr)$);
    \draw (m1) -- ($(m1)-(yr)$);
    \draw (m2) -- ($(m2)-(yr)$);
    \draw (m3) -- ($(m3)-(yr)$);
    \draw (m4) -- ($(m4)-(yr)$);
    
    \draw (j0) -- ($(j0)-(xr)$);
    \draw (j1) -- ($(j1)-(xr)$);
    \draw (j2) -- ($(j2)-(xr)$);
    \draw (j3) -- ($(j3)-(xr)$);
    \draw (j4) -- ($(j4)-(xr)$);
    \draw (j5) -- ($(j5)-(xr)$);
    \draw (j6) -- ($(j6)-(xr)$);
    
    \node at ($(m0)-2.5*(yr)$) {$\scriptstyle 0$};
    \node at ($(m1)-2.5*(yr)$) {$\scriptstyle 1$};
    \node at ($(m2)-2.5*(yr)$) {$\scriptstyle 2$};
    \node at ($(m3)-2.5*(yr)$) {$\scriptstyle 3$};
    \node at ($(m4)-2.5*(yr)$) {$\scriptstyle 4$};

    \node at ($(j0)-2.5*(xr)$) {$\scriptstyle 0$};
    \node at ($(j1)-2.5*(xr)$) {$\scriptstyle 1$};
    \node at ($(j2)-2.5*(xr)$) {$\scriptstyle 2$};
    \node at ($(j3)-2.5*(xr)$) {$\scriptstyle 3$};
    \node at ($(j4)-2.5*(xr)$) {$\scriptstyle 4$};
    \node at ($(j5)-2.5*(xr)$) {$\scriptstyle 5$};
    \node at ($(j6)-2.5*(xr)$) {$\scriptstyle 6$};

    \draw[color=red, thick, dashed]    ( 0.0, 2.0)--( 2.3, 6.6);
    \draw[color=orange, thick, dashed] ( 0.0, 0.0)--( 3.3, 6.6);
    \draw[color=yellow, thick, dashed] ( 1.0, 0.0)--( 4.3, 6.6);
    \draw[color=green, thick, dashed]  ( 2.0, 0.0)--( 4.3, 4.6);
    \draw[color=blue, thick, dashed]   ( 3.0, 0.0)--( 4.3, 2.6);
    \draw[color=violet, thick, dashed] ( 4.0, 0.0)--( 4.3, 0.6);

    \filldraw[color=red]    (0,2) circle (2.5pt);
    \filldraw[color=red]    (1,4) circle (2.5pt);
    \filldraw[color=red]    (2,6) circle (2.5pt);

    \filldraw[color=orange] (0,0) circle (2.5pt);
    \filldraw[color=orange] (1,2) circle (2.5pt);
    \filldraw[color=orange] (2,4) circle (2.5pt);
    \filldraw[color=orange] (3,6) circle (2.5pt);
    
    \filldraw[color=yellow] (1,0) circle (2.5pt);
    \filldraw[color=yellow] (2,2) circle (2.5pt);
    \filldraw[color=yellow] (3,4) circle (2.5pt);
    \filldraw[color=yellow] (4,6) circle (2.5pt);
    
    \filldraw[color=green]  (2,0) circle (2.5pt);
    \filldraw[color=green]  (3,2) circle (2.5pt);
    \filldraw[color=green]  (4,4) circle (2.5pt);

    \filldraw[color=blue]   (3,0) circle (2.5pt);
    \filldraw[color=blue]   (4,2) circle (2.5pt);

    \filldraw[color=violet] (4,0) circle (2.5pt);
\end{tikzpicture}
\\ \bigskip
\begin{tikzpicture}
[scale=0.7]
    \coordinate (m0) at (0,0);
    \coordinate (m1) at ($(  5.00^2/55, 0.0)$);
    \coordinate (m2) at ($( 10.00^2/55, 0.0)$);
    \coordinate (m3) at ($( 15.00^2/55, 0.0)$);
    \coordinate (m4) at ($( 20.00^2/55, 0.0)$);
    \coordinate (m5) at ($( 25.00^2/55, 0.0)$);

    \coordinate (j0) at ( 0.0, 0.0);
    \coordinate (j1) at ( 0.0, 1.0);
    \coordinate (j2) at ( 0.0, 2.0);
    \coordinate (j3) at ( 0.0, 3.0);
    \coordinate (j4) at ( 0.0, 4.0);
    \coordinate (j5) at ( 0.0, 5.0);
    \coordinate (j6) at ( 0.0, 6.0);

    \node[anchor=mid] at (13.5, 0.0)
         {$\displaystyle\frac{m^2}{\text{GeV}^2}$};
    \node at ( 0.0, 7.0) {$j$};
    
    \draw[->] ( 0.0, 0.0)--(12.5, 0.0);
    \draw[->] ( 0.0, 0.0)--( 0.0, 6.5);

    \coordinate (xr) at ( 0.2, 0.0);
    \coordinate (yr) at ( 0.0, 0.2);

    \draw (m0) -- ($(m0)-(yr)$);
    \draw (m1) -- ($(m1)-(yr)$);
    \draw (m2) -- ($(m2)-(yr)$);
    \draw (m3) -- ($(m3)-(yr)$);
    \draw (m4) -- ($(m4)-(yr)$);
    \draw (m5) -- ($(m5)-(yr)$);
    
    \draw (j0) -- ($(j0)-(xr)$);
    \draw (j1) -- ($(j1)-(xr)$);
    \draw (j2) -- ($(j2)-(xr)$);
    \draw (j3) -- ($(j3)-(xr)$);
    \draw (j4) -- ($(j4)-(xr)$);
    \draw (j5) -- ($(j5)-(xr)$);
    \draw (j6) -- ($(j6)-(xr)$);
    
    \node at ($(m0)-2.5*(yr)$) {};
    \node at ($(m1)-2.5*(yr)$) {$\scriptstyle {}^{\phantom{2}}0.5^2$};
    \node at ($(m2)-2.5*(yr)$) {$\scriptstyle {}^{\phantom{2}}1.0^2$};
    \node at ($(m3)-2.5*(yr)$) {$\scriptstyle {}^{\phantom{2}}1.5^2$};
    \node at ($(m4)-2.5*(yr)$) {$\scriptstyle {}^{\phantom{2}}2.0^2$};
    \node at ($(m5)-2.5*(yr)$) {$\scriptstyle {}^{\phantom{2}}2.5^2$};

    \node at ($(j0)-2.5*(xr)$) {$\scriptstyle 0$};
    \node at ($(j1)-2.5*(xr)$) {$\scriptstyle 1$};
    \node at ($(j2)-2.5*(xr)$) {$\scriptstyle 2$};
    \node at ($(j3)-2.5*(xr)$) {$\scriptstyle 3$};
    \node at ($(j4)-2.5*(xr)$) {$\scriptstyle 4$};
    \node at ($(j5)-2.5*(xr)$) {$\scriptstyle 5$};
    \node at ($(j6)-2.5*(xr)$) {$\scriptstyle 6$};

    \coordinate (rho-770)   at ($( 7.75^2/55, 1)$);
    \coordinate (rho3-1690) at ($(16.89^2/55, 3)$);
    \coordinate (rho5-2350) at ($(23.30^2/55, 5)$);

    \coordinate (f2-1270)   at ($(12.75^2/55, 2)$);
    \coordinate (f4-2050)   at ($(20.18^2/55, 4)$);  
    \coordinate (f6-2510)   at ($(24.70^2/55, 6)$);

    \filldraw[color=red] (rho-770)   circle (2.5pt);
    \filldraw[color=red] (rho3-1690) circle (2.5pt);
    \filldraw[color=red] (rho5-2350) circle (2.5pt);

    \filldraw[color=red] (f2-1270)   circle (2.5pt);
    \filldraw[color=red] (f4-2050)   circle (2.5pt);
    \filldraw[color=red] (f6-2510)   circle (2.5pt);

    \node[above, color=red] at ($(rho-770)  +(0,0.2)$)
        {$\scriptstyle \rho(770)$};
    \node[above, color=red] at ($(f2-1270)  +(0,0.2)$)
        {$\scriptstyle f_2(1270)$};
    \node[above, color=red] at ($(rho3-1690)+(0,0.2)$)
        {$\scriptstyle \rho_3(1690)$};
    \node[above, color=red] at ($(f4-2050)  +(0,0.2)$)
        {$\scriptstyle f_4(2050)$};
    \node[above, color=red] at ($(rho5-2350)+(0,0.2)$)
        {$\scriptstyle \rho_5(2350)$};
    \node[above, color=red] at ($(f6-2510)  +(0,0.2)$)
        {$\scriptstyle f_6(2510)$};

\end{tikzpicture}
\hfill

\caption{Spin $j$ vs.\ mass-squared $m^2$ of the open superstring spectrum (top-left), the closed superstring spectrum (top-right), and the leading observed $\rho$~and~$f$ mesons~\cite{ParticleDataGroup:2024cfk} (bottom) in mass-squared units of $(\alpha')^{-1}$, $(\tfrac{1}{4}\alpha')^{-1}$, and $\text{GeV}^2$, respectively. In each theory, the spectrum can be organized into Regge trajectories with ${j \propto m^2}$. For simplicity, we do not plot the daughter $\rho$~and~$f$ states although many have been observed~\cite{ParticleDataGroup:2024cfk}.}

\label{fig:regge}

\end{figure}
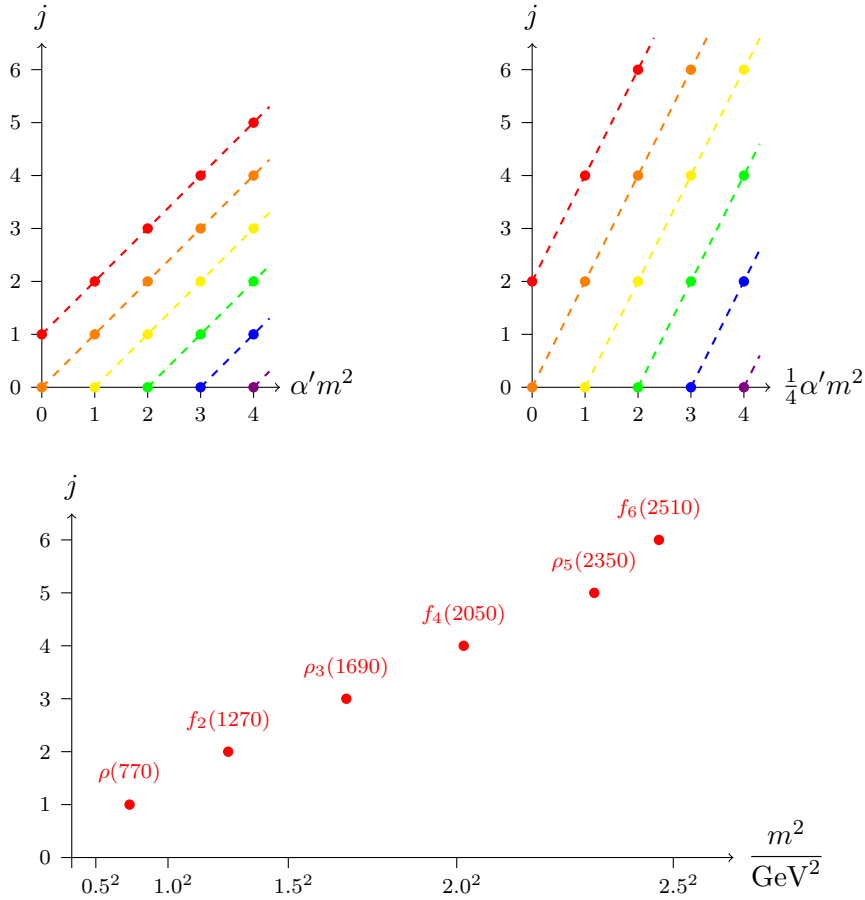

There is a deep relationship between the consistency of a low-energy effective theory and the spectrum of massive spinning states in its UV completion. This relationship is beautifully exhibited in both string theory and real-world QCD. In \autoref{fig:regge}, we present the Regge trajectories (spin vs.\ mass-squared) of the open superstring spectrum, the closed superstring spectrum, and a subset of light mesons in QCD. In each case, the states are organized into linear (or approximately linear) Regge trajectories. This observation dates back to the old days of the S-matrix bootstrap program~\cite{Cappelli:2012cto, chew1966analytic, Eden:1966dnq} and can be extended to the whole zoo observed of mesons~\cite{ParticleDataGroup:2024cfk, Chen:2021kfw}.

The organization of states into linear Regge trajectories hints at some underlying principle. Surely, states cannot appear willy-nilly with any mass and spin! Any such principle falls within the scope of the bootstrap paradigm and can be phrased as a question:
\begin{quote}
\textit{Do the S-matrix bootstrap assumptions of unitarity, causality, and locality imply bounds on the masses and spins of states in a relativistic quantum theory?}
\end{quote}
To this end, it was recently observed numerically~\cite{Berman:2024wyt} and then subsequently proven analytically~\cite{Berman:2024kdh} that there exist bounds on the spins and masses of the states exchanged in any weakly-coupled theory with massless states and planar scattering amplitudes (such as large-$N$ gauge theories). In~\cite{Albert:2024yap}, a similar numerical observation was made regarding the maximum spins exchanged in super-graviton amplitudes. We name these bounds the \textbf{``Sequential Spin Constraints" (SSC)} and \textbf{``Sequential Mass Constraints" (SMC)}. The SSC and SMC follow from the usual S\nobreakdash-matrix bootstrap assumptions and can be proved by studying the dispersion relations for various low-energy effective couplings.

\newpage

For each case considered in this paper, the SSC and SMC can be summarized in terms of the masses $\mu_j$ of the lightest spin-$j$ particles. Modulo some technical details,
\begin{align}
\label{eq:SSCSMC}
\boxed{
    \quad
    \underbrace{
    \mu_{j\vphantom{j'}} < \mu_{j'}
    \vphantom{\Bigg|}
        }_{\text{SSC}}
    \qquad
    \text{and}
    \qquad
    \underbrace{
    \bigg(
    \frac{ \mu_{j''} }{ \mu_{j\phantom{''}} }
    \bigg)^{j'-j}
    \leq
    \bigg(
    \frac{ \mu_{j'} }{ \mu_{j\phantom{'}} }
    \bigg)^{j''-j}
    \vphantom{\Bigg|}
        }_{\text{SMC}}
    \qquad
    \text{for spins }
    {j < j' < j''}
    \, .
    \quad
}
\end{align}
In other words, the SSC requires the lightest spin-$j$ state to be lighter than the lightest spin-$j'$ state. The SMC requires the mass of this spin-$j$ state to be smaller than a non-linear function of the masses of lower-spin states. The stringy Regge trajectories depicted in~\autoref{fig:regge} saturate the SSC, but the consequences of the SMC are less obvious.

In~\cite{Berman:2024kdh}, the SSC and SMC were applied to QCD in the large-$N$ approximation and the chiral ${N_F = 2}$ limit (with massless $ud$ quarks), in which case the scattering amplitudes are planar and the pions are massless. In this approximation, only $\rho$ and~$f$ mesons are exchanged in the two-to-two scattering of pions~\cite{Albert:2022oes}. The experimentally observed $\rho$~and~$f$ mesons with spins ${j=1,2,3,4,5,6}$ satisfy both the SSC and SMC bounds~\cite{ParticleDataGroup:2024cfk, Berman:2024kdh}. Moreover, the SSC and SMC predict bounds for the mass of the yet-to-be-discovered $\rho_7$ meson~\cite{Berman:2024kdh}. The SSC requires the mass of this spin-$7$ meson to be greater than the mass of the lightest spin-$6$ meson exchanged in the four-pion amplitude: ${m_{f_6} < m_{\rho_7}}$. The SMC requires the mass of this spin-$7$ meson to be less than or equal to the squared mass of the lightest spin-$6$ meson divided by the mass of the lightest spin-$5$ meson: $\smash{m_{\rho_7} \leq m_{f_6}^2 / m_{\rho_5}}$. Using the PDG values~\cite{ParticleDataGroup:2024cfk} for $m_{\rho_5}$ and $m_{f_6}$, we find
\begin{align}
\label{eq:rhobound}
    2470 \text{ MeV}
    <
    m_{\rho_7}
    \leq
    2618 \text{ MeV}
    \, .
\end{align}
These bounds provide a testable prediction for QCD at an experimentally accessible energy!

We emphasize that this analysis is strictly true only for QCD with a large number of colors~$N$, with ${N_F=2}$ massless quarks, with massless pions, with stable mesons, and at weak-coupling. This set-up appears throughout the modern bootstrap literature~\mbox{\cite{Albert:2022oes, Albert:2023jtd, Albert:2023seb, Fernandez:2022kzi, Li:2023qzs}}, and the validity of these assumptions are discussed in~\cite{Albert:2022oes}. However, real-world QCD violates these assumptions with $N=3$ and massive quarks, so the bound~\eqref{eq:rhobound} on the mass of the $\rho_7$ meson is only approximate.

In this paper, we extend the analytic results of~\cite{Berman:2024kdh} beyond the case of planar four-point amplitudes. We consider any theory with massless scalar particles, with or without gravity, and with as few simplifying assumptions as possible. We also discuss the extension of our results to super-gluon and super-graviton amplitudes.

Our approach is mathematical, and we phrase the SSC and SMC, summarized in~\eqref{eq:SSCSMC}, as theorems. The proof here also differs from that of~\cite{Berman:2024kdh}, which used the positivity of certain Hankel matrices at a crucial intermediate step. We use Jensen's inequality~\cite{10.5555/26851} at the same juncture, which simplifies the argument. By taking a methodical approach, we identify the key steps of our proof, which we hope can be further generalized to a wider class of theories, such as those with non-supersymmetric photons and gravitons, with massive external states, or without the requirement of weak-coupling.

Our proof makes use of fixed-$u$ dispersion relations, an old tool~\cite{Chew:1957tf, Mandelstam:1958xc, Jin:1964zza} with widespread use in the modern S-matrix bootstrap literature~\cite{Arkani-Hamed:2020blm, Caron-Huot:2020cmc}. Using the bootstrap assumptions, we write dispersion relations for the Wilson coefficients of a generic four-point scalar amplitude. The dispersion relations exhibit exactly how IR observables, i.e.\ the Wilson coefficients, depend on the UV spectrum of the theory. These dispersion relations can be organized into a moment problem~\cite{Bellazzini:2020cot,Arkani-Hamed:2020blm,Chiang:2021ziz}, which implies that the Wilson coefficients obey an infinite number of non-linear inequalities. Crossing symmetry further implies that the Wilson coefficients obey an infinite number of linear equalities. By carefully combining these relations and taking a well-controlled infinite limit in the order~$k$ of the derivative expansion, we arrive at the SSC and SMC. Because numerical bootstrap studies truncate the problem at some fixed $k_\text{max}$, hints of these analytic results can be seen by plotting numerical results to large $k_\text{max}$~\cite{Berman:2024wyt, Berman:2024eid, Albert:2024yap}.

In the remainder of this section, we precisely state our conventions and bootstrap assumptions in order to form a strong mathematical basis for the proofs which follow.

\subsection{Conventions}

In this paper, we consider the two-to-two scattering process for four identical massless scalar particles in flat spacetime with an arbitrary number $d \geq 3$ of spacetime dimensions. We use the mostly-plus signature $\eta_{\mu \nu} = \operatorname{diag}(-1,1,\dots,1)$. We take all momenta incoming so that the statement of momentum conservation is ${p_1 + p_2 + p_3 + p_4 = 0}$. The Mandelstam variables ${s_{ij} = -(p_i+p_j)^2}$ for this scattering process are
\begin{alignat}{3}
\label{eq:stu}
    s
    &=
    s_{12}
    =
    s_{34}
    =
    \phantom{-} 4 E^2
    &&
    {} \geq 0
    \, ,
\no \\
    t
    &=
    s_{13}
    =
    s_{24}
    =
    -2 E^2 (1+\cos\theta)    
    &&
    {} \leq 0
    \, ,
\no \\
    u
    &=
    s_{14}
    =
    s_{23}
    =
    -2 E^2 (1-\cos\theta)
    &&
    {} \leq 0
    \, .
\end{alignat}
Here~$E$ and~$\theta$ are the center-of-mass energy and scattering angle, respectively, and the inequalities refer to the physical scattering regime with real~$s_{ij}$.

The three Mandelstam variables satisfy the on-shell relation~${s+t+u=0}$, leaving two independent variables. We work with the variables $s$ and~$u$. This choice corresponds to the canonical cyclic ordering $1234$ of the external states.

Any scattering amplitude for four massless scalar particles is a function $A(s,u)$ of the variables $s$ and~$u$ with mass dimension ${[A(s,u)]=4-d}$. More precisely, amplitudes are distributions rather than functions, but we can safely ignore this technicality. As usual, we assume that the physical amplitude with real ${s \geq -u \geq 0}$ is the boundary value of an analytic function of complex $s$ and~$u$. Any non-analyticities in $s$ and~$u$, such as simple poles or branch cuts, are dictated by unitarity and the physical content of the theory. For fixed $u < 0$, these non-analyticities will only occur at real $s$, so it is useful to study the discontinuity of $A(s,u)$ across the real $s$-axis.

For any function $f(s)$ of complex $s$, we identify its imaginary part $\Im f(s)$ with real~$s$ as the discontinuity across the real $s$-axis:
\begin{align}
    \Im f(s) 
    = 
    \frac{1}{2i}
    \lim_{\eps \to 0^+}
    \Big(
    f(s+i\eps) - f(s-i\eps)
    \Big)
    \, .
\end{align}
This definition encodes the Feynman $i\eps$ prescription so that the imaginary part of a propagator is proportional to the Dirac delta function:
\begin{align}
    \Im \frac{1}{s-m^2} 
    = 
    - \pi \, \delta(s-m^2)
    \, .
\end{align}
Hence, a delta function with support at $s=m^2$ in the imaginary part of an amplitude indicates the tree-level exchange of a particle with mass $m$. Similarly, the logarithm ${\ln(s-m^2)}$ arises in an amplitude when a particle of mass $m$ flows around a one-loop Feynman diagram. This logarithm contributes a constant term with support on $s \geq m^2$ to the imaginary part of the amplitude:
\begin{align}
    \Im \ln(s-m^2) 
    =
    - \pi \, \Theta(s-m^2)
    \, ,
\end{align}
where $\Theta(x)$ is the Heaviside theta function.

Throughout this paper, we consider two cases of four-point amplitudes with two distinct analytic structures: $su$-symmetric and $stu$-symmetric amplitudes, which we denote by $A^{(su)}(s,u)$ and $A^{(stu)}(s,u)$, respectively. We use $A(s,u)$ whenever the distinction is immaterial. The terms ``crossing symmetry" and ``permutation symmetry" are often used in the literature to refer to $su$- and $stu$-symmetry, respectively, but in this paper we exclusively use the more precise terminology of $su$- and $stu$-symmetry, reserving crossing symmetry for the general property of invariance under label changes.

The $su$-symmetric amplitudes arise as amplitudes for a particular cyclic ordering of the external states stripped of their color or flavor structure~\cite{Dixon:1996wi, Elvang:2013cua}, and $A^{(su)}(s,u)$ corresponds to the canonical cyclic ordering $1234$. These amplitudes are the relevant object to study if, for example, our external scalars transform in the adjoint representation of $\mathrm{SU}(N)$ with large ${N \gg 1}$. In this case, the full scattering amplitude for any process decomposes into a sum of these color- or flavor- stripped-amplitudes multiplied by single-trace group theoretical factors, and we can study these stripped-amplitudes without specifying the values of the color or flavor indices of the external states. Multi-trace contributions (such as those containing $t$\nobreak-channel cuts and poles) are suppressed by $1/N$, so the leading contribution in the large-$N$ limit has only $s$-channel and $u$-channel cuts and poles.

The $stu$-symmetric amplitudes are simply the full four-point scattering amplitudes for four identical external states. Particular examples include dilatons or Goldstone bosons. In general, the scalars may have color or flavor indices, but the values of these indices must be identical for the amplitude to have full $stu$-symmetry.

\subsection{Bootstrap assumptions}

We further assume that our amplitudes obey a number of technical properties, all of which are standard assumptions in the modern S-matrix bootstrap literature. For the sake of precision, we state each property as clearly as possible. The logical interdependence and technical details of these assumptions are discussed in~\cite{Caron-Huot:2016icg, Correia:2020xtr, Haring:2022cyf} and the references therein.
\begin{enumerate}

\item
\label{prop:crossing}
\textbf{Crossing symmetry}

Our amplitudes are crossing symmetric, i.e.\ invariant under permutations of the external particle labels (and thus under permutations of the Mandelstam variables). For $A^{(su)}(s,u)$, we only consider permutations which respect the cyclic ordering of the external states, and for $A^{(stu)}(s,u)$, we consider all permutations. Hence,
\begin{align}
\label{eq:crossing-su}
      A^{(su)}(s,u)
    = A^{(su)}(u,s)
\end{align}
and
\begin{align}
\label{eq:crossing-stu}
      A^{(stu)}(s,u)
    &
    = A^{(stu)}(u,t)
    = A^{(stu)}(t,s)
\no \\
    = A^{(stu)}(u,s)
    &
    = A^{(stu)}(t,u)
    = A^{(stu)}(s,u)
\end{align}
with $t=-(s+u)$. Although crossing symmetry is a simple property to state, its precise formulation and proof for generic non-perturbative amplitudes is an old problem with many recent developments~\cite{Mizera:2021ujs, Mizera:2021fap, Caron-Huot:2023ikn}.

\item
\label{prop:bound}
\textbf{Polynomial boundedness}

Our amplitudes are polynomially-bounded in the high-energy Regge limit of large~$|s|$ (away from any non-analyticities on the real $s$-axis) and fixed ${u<0}$. In other words, there exists a smallest integer $\alpha$ such that
\begin{align}
\label{eq:bound}
    \lim_{\substack{ |s| \to \infty \\
                     \mathclap{\text{fixed } u<0}}}
    \frac{A(s,u)}{s^\alpha} 
    = 0
    \, .
\end{align}
This bound is related to the famous Froissart-Martin bound~\cite{Froissart:1961ux, Martin:1962rt}, and the standard behavior for well-behaved theories is ${\alpha = 2}$~\cite{Arkani-Hamed:2020blm, Correia:2020xtr, Haring:2022cyf}. Stronger bounds can be justified under various assumptions, but to be as general as possible, we leave the integer~$\alpha$ unspecified. The notation $\alpha$ is chosen in analogy with the notation of the old string theory and S\nobreakdash-matrix literature~\cite{Cappelli:2012cto}. Its value may differ between $A^{(su)}(s,u)$ and $A^{(stu)}(s,u)$ and between particular theories. Non-polynomial bounds were recently considered in~\cite{Buoninfante:2024ibt}, but we restrict ourselves to the more physically motivated polynomial bound.

\item
\label{prop:partial}
\textbf{Partial wave decomposition}

For physical kinematics (i.e.\ for real ${s \geq -u > 0}$), our amplitudes can be written in a basis of partial waves as follows:
\begin{align}
\label{eq:partial}
    A(s,u)
    &=
    \sum_{j = 0}^\infty
    n_j^{(d)}
    a_j^{\mathstrut}(s) \,
    \mcP_j^{(d)} \Big( 1 + \frac{2u}{s} \Big)
    \, ,
\end{align}
where the $d$-dimensional spin-$j$ Gegenbauer polynomial is defined by
\begin{align}
    \mcP_j^{(d)}(x)
    &=
    {}_2 F_1
    \Big(
    {-j}, \, 
    j+d-3; \,
    \tfrac{1}{2} (d-2); \,
    \tfrac{1}{2} (1-x) 
    \Big)
    \, ,
\end{align}
the normalization factor is defined by
\begin{align}
\label{eq:Gegn}
    n_j^{(d)}
    &=
    \frac{ (4\pi)^{\frac{d}{2}} (2j+d-3) \, \Gamma(j+d-3) }
         { \pi \, \Gamma(\frac{d-2}{2}) \, \Gamma(j+1) }
    \, ,
\end{align}
and $a_j(s)$ is the spin-$j$ partial wave.

In the $su$-symmetric case, both even and odd spins can contribute to the partial wave expansion~\eqref{eq:partial}. In the $stu$-symmetric case, only even spins will contribute to~\eqref{eq:partial} because the $ut$-symmetry relation,
$$
    A^{(stu)}(s,u) = A^{(stu)}(s,-s{-u})
    \, ,
$$
and the parity property of the Gegenbauer polynomials,
$$
    \mcP_{j}^{(d)}(-x)
    =
    (-)^{j} \, \mcP_{j}^{(d)}(x)
    \, ,
$$
imply that all the odd partial waves vanish.

\item
\label{prop:unitarity}
\textbf{Unitarity (Positivity)}

Our amplitudes obey unitarity, which can be phrased in terms of an inequality satisfied by each spin-$j$ partial wave $a_j(s)$ for physical values of ${s \geq 0}$:
\begin{align}
\label{eq:unitarity}
    \big|
    1 + i s^{\frac{d-4}{2}} a_j(s)
    \big|^2
    \leq 1
    \, .
\end{align}
Unitarity implies that the spin-$j$ spectral density functions $\rho_j(s)$, which we define by
\begin{align}
\label{eq:rho}
    \rho_j(s)
    =
    \frac{1}{\pi} \, n_j^{(d)}
    s^{\frac{d-4}{2}} \Im a_j(s)
    \, ,
\end{align}
satisfy
\begin{align}
\label{eq:positivity}
    \rho_j(s) \geq 0
    \, .
\end{align}
This property is called the positivity of the spectral density.

Positivity~\eqref{eq:positivity} rather than full non-linear unitarity relation~\eqref{eq:unitarity} is the appropriate assumption for weakly-coupled theories. At weak-coupling, the amplitude $A(s,u)$ and its partial waves $a_j(s)$ are parametrically small so that~\eqref{eq:unitarity} reduces to~\eqref{eq:positivity} at leading order.

In any case, these inequalities should be read as distributional inequalities rather than point-wise inequalities to allow for poles, delta functions, etc.\ along the positive $s$\nobreakdash-axis. For example,~\eqref{eq:positivity} should technically be read as $\int_0^\infty \mathrm{d}s \, f(s) \, \rho_j(s) \geq 0$ for any positive test function ${f(s) \geq 0}$.

\item
\label{prop:analyticity}
\textbf{Weak-coupling analyticity}

Our amplitudes have a particular analytic structure which we call weak-coupling analyticity, meaning they describe theories which are weakly-coupled below a cut-off scale $\Lambda$. This cut-off can be finite or infinite. If $\Lambda = \infty$, then the theory is weakly-coupled at all energies, and the amplitude is a meromorphic function of the Mandelstam variables.

At energies $E \ll \Lambda$, the amplitudes are well-approximated by the tree-level exchanges of a discrete set of intermediate particle states with masses ${m_n < \Lambda}$, which correspond to simple poles in the Mandelstam variables. Without loss of generality, we allow for either a finite or an infinite set of masses and order them by ${0 = m_0 < m_1 < \cdots < \Lambda}$. The weakly-coupled spectrum may exhibit an accumulation point, in which case there are an infinite set of masses and a finite cut-off with $\lim_{n \to \infty} m_n = \Lambda$.

At energies $E \geq \Lambda$, we are agnostic to the physics and the strength of the interactions, allowing for loop effects and branch cuts in addition to simple poles. Specifically, we assume that the $su$- and $stu$-symmetric amplitudes have the following analytic structure, depicted in~\autoref{fig:analyticity}.

At fixed $u < 0$, $A^{(su)}(s,u)$ is an analytic function of $s \in \bbC$ with only the following non-analyticities on the real $s$-axis:
\begin{itemize}
    \item simple poles in $s$ at each $s = m_n^2$
    \item possible branch cuts or simple poles in $s$ along $s \in [\Lambda^2, \infty)$
\end{itemize}
At fixed $u < 0$, $A^{(stu)}(s,u)$ is an analytic function of $s \in \bbC$ with only the following non-analyticities on the real $s$-axis:
\begin{itemize}
    \item simple poles in $s$ at each $s = m_n^2$ and $s=-u-m_n^2$
    \item possible branch cuts or simple poles in $s$ along $s \in [\Lambda^2, \infty)$ and $s \in (-\infty, -u-\Lambda^2]$
\end{itemize}

\begin{figure}
\centering

\begin{tikzpicture}
    \clip (-5,-1) rectangle (5,3);

    \draw[fill=gray!20, color=gray!20]
         (-5.0,-1.0) --
         ( 5.0,-1.0) --
         ( 5.0, 3.0) --
         (-5.0, 3.0) -- cycle;
    
    \coordinate (m0) at ( 0.0, 0.0);
    \coordinate (m1) at ( 1.4, 0.0);
    \coordinate (mn) at ( 2.7, 0.0);
    \coordinate (mL) at ( 4.0, 0.0);

    \coordinate (u)  at (-0.6, 0.0);
    \coordinate (u0) at ($-1*(u)-(m0)$);

    \node at ( 4.5, 0.5) {$\scriptstyle\Re(s)$};
    \node at (-0.6, 2.6) {$\scriptstyle\Im(s)$};

    \coordinate (xr) at ( 0.3, 0.0);
    \coordinate (yr) at ( 0.0, 0.2);
    
    \draw     (-5.0, 0.0)   -- ($(m1)+(xr)$);
    \draw     ($(mn)-(xr)$) -- ($(mn)+(xr)$);
    \draw[->] ($(mL)-(xr)$) -- ( 5.0, 0.0);
    
    \draw[->] ( 0.0, 0.0)   -- ( 0.0, 3.0);

    \draw (m0) -- ($(m0)-(yr)$);
    \draw (m1) -- ($(m1)-(yr)$);
    \draw (mn) -- ($(mn)-(yr)$);
    \draw (mL) -- ($(mL)-(yr)$);
    \draw (u0) -- ($(u0)-(yr)$);
    
    \node at ($(m0)-2.5*(yr)$) {$\scriptstyle 0 \mathstrut$};
    \node at ($(m1)-2.5*(yr)$) {$\scriptstyle ^{\phantom{2}} m_1^2$};
    \node at ($(mn)-2.5*(yr)$) {$\scriptstyle ^{\phantom{2}} m_n^2$};
    \node at ($(mL)-2.5*(yr)$) {$\scriptstyle ^{\phantom{2}} \Lambda_{\phantom{n}}^2$};

    \node at ($(u0)-2.5*(yr)$) {$\scriptstyle -u \phantom{-}$};

    \node at ($0.5*(mn)+0.5*(m1)$) {$\cdots$};
    \node at ($0.5*(mL)+0.5*(mn)$) {$\cdots$};

    \filldraw[color=red] (m0) circle (2pt);
    \filldraw[color=red] (m1) circle (2pt);
    \filldraw[color=red] (mn) circle (2pt);
    \filldraw[color=red] (mL) circle (1pt);

    \draw[color=red, very thick, decorate,
          decoration={zigzag, amplitude=2pt, segment length=5pt}]
        (mL) -- ( 4.95, 0.0);   

    \node[fill=white] at (3.8,2.5) {$A^{(su)}(s,u)$};
    
\end{tikzpicture}
\\ \medskip
\begin{tikzpicture}
    \clip (-5,-1) rectangle (5,3);

    \draw[fill=gray!20, color=gray!20]
         (-5.0,-1.0) --
         ( 5.0,-1.0) --
         ( 5.0, 3.0) --
         (-5.0, 3.0) -- cycle;
    
    \coordinate (m0) at ( 0.0, 0.0);
    \coordinate (m1) at ( 1.4, 0.0);
    \coordinate (mn) at ( 2.7, 0.0);
    \coordinate (mL) at ( 4.0, 0.0);

    \coordinate (u)  at (-0.6, 0.0);

    \coordinate (u0) at ($-1*(u)-(m0)$);
    \coordinate (u1) at ($-1*(u)-(m1)$);
    \coordinate (un) at ($-1*(u)-(mn)$);
    \coordinate (uL) at ($-1*(u)-(mL)$);

    \node at ( 4.5, 0.5) {$\scriptstyle\Re(s)$};
    \node at (-0.6, 2.6) {$\scriptstyle\Im(s)$};

    \coordinate (xr) at ( 0.3, 0.0);
    \coordinate (yr) at ( 0.0, 0.2);
    
    \draw     (-5.0, 0.0)   -- ($(uL)+(xr)$);
    \draw     ($(un)-(xr)$) -- ($(un)+(xr)$);
    \draw     ($(u1)-(xr)$) -- ($(m1)+(xr)$);
    \draw     ($(mn)-(xr)$) -- ($(mn)+(xr)$);
    \draw[->] ($(mL)-(xr)$) -- ( 5.0, 0.0);
    
    \draw[->] ( 0.0, 0.0)   -- ( 0.0, 3.0);

    \draw (m0) -- ($(m0)-(yr)$);
    \draw (m1) -- ($(m1)-(yr)$);
    \draw (mn) -- ($(mn)-(yr)$);
    \draw (mL) -- ($(mL)-(yr)$);

    \draw (u0) -- ($(u0)-(yr)$);
    \draw (u1) -- ($(u1)-(yr)$);
    \draw (un) -- ($(un)-(yr)$);
    \draw (uL) -- ($(uL)-(yr)$);
    
    \node at ($(m0)-2.5*(yr)$) {$\scriptstyle 0 \mathstrut$};
    \node at ($(m1)-2.5*(yr)$) {$\scriptstyle ^{\phantom{2}} m_1^2$};
    \node at ($(mn)-2.5*(yr)$) {$\scriptstyle ^{\phantom{2}} m_n^2$};
    \node at ($(mL)-2.5*(yr)$) {$\scriptstyle ^{\phantom{2}} \Lambda_{\phantom{n}}^2$};

    \node at ($(u0)-2.5*(yr)$) {$\scriptstyle -u \phantom{-}$};
    \node at ($(u1)-2.5*(yr)$) {$\scriptstyle -u-m_1^2 \,\,$};
    \node at ($(un)-2.5*(yr)$) {$\scriptstyle -u-m_n^2 \,\,$};
    \node at ($(uL)-2.5*(yr)$) {$\scriptstyle -u-\Lambda_{\phantom{n}}^2 \,\,$};

    \node at ($0.5*(mn)+0.5*(m1)$) {$\cdots$};
    \node at ($0.5*(mL)+0.5*(mn)$) {$\cdots$};

    \node at ($0.5*(un)+0.5*(u1)$) {$\cdots$};
    \node at ($0.5*(uL)+0.5*(un)$) {$\cdots$};

    \filldraw[color=red] (m0) circle (2pt);
    \filldraw[color=red] (m1) circle (2pt);
    \filldraw[color=red] (mn) circle (2pt);
    \filldraw[color=red] (mL) circle (1pt);

    \filldraw[color=red] (u0) circle (2pt);
    \filldraw[color=red] (u1) circle (2pt);
    \filldraw[color=red] (un) circle (2pt);
    \filldraw[color=red] (uL) circle (1pt);

    \draw[color=red, very thick, decorate,
          decoration={zigzag, amplitude=2pt, segment length=5pt}]
        (mL) -- ( 4.95, 0.0);   
        
    \draw[color=red, very thick, decorate,
          decoration={zigzag, amplitude=2pt, segment length=5pt}]
        (uL) -- (-5.0, 0.0);   

    \node[fill=white] at (3.8,2.5) {$A^{(stu)}(s,u)$};
    
\end{tikzpicture}

\caption{The analytic structure of the amplitudes $A^{(su)}(s,u)$ and $A^{(stu)}(s,u)$ for complex $s$ and fixed $u < 0$ (assuming weak coupling analyticity). Isolated red dots indicate simple poles in $s$, and the red zig-zag lines indicate poles or branch cuts.}
\label{fig:analyticity}

\end{figure}
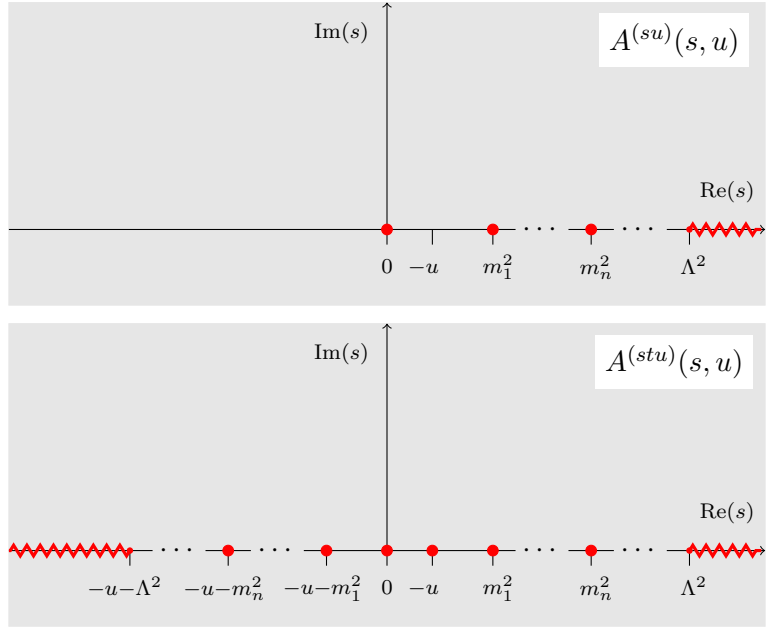

\item
\label{prop:lowenergy}
\textbf{Low-energy expansion}

Our amplitudes admit a low-energy expansion near $s = u = 0$ of the form
\begin{align}
\label{eq:lowenergy}
    A(s,u) 
    &=
    A_{\text{massless}}(s,u)
    +
    \sum_{k = 0}^\infty
    \sum_{q = 0}^k
    a_{k,q} \,
    s^{k-q} u^q
    \, ,
\end{align}
where $A_{\text{massless}}(s,u)$ contains the simple poles from the tree-level exchange of massless states at $s=0$ and $u=0$ (in the $su$-symmetric case) or at $s=0$, $u=0$, and $s+u=0$ (in the $stu$-symmetric case). This expression is valid for $|s|, |u| < m_1^2$ in the $su$-symmetric case and for $|s|, |u|, |s+u| < m_1^2$ in the $stu$-symmetric case.

The $a_{k,q}$ are called Wilson coefficients and have mass dimension $[a_{k,q}]=4-d-2k$. The set of $a_{k,q}$ with fixed $k$ correspond to operators of the form $\partial^{2k} \phi^4$ in an effective Lagrangian (where $\phi$ is the massless scalar field). Hence, the values of the Wilson coefficients parametrize the space of EFTs with a given massless spectrum.

At weak-coupling, $A(s,u)$ and its Wilson coefficients are parametrically small, so modern bootstrap methods usually study ratios of Wilson coefficients (or other ratios of observables) in which the overall small parameter cancels. Formally, this small parameter can be factored out of $A(s,u)$ without changing any of our assumptions, so we can formally study individual Wilson coefficients (and other parametrically small observables) without always writing ratios.

\item
\label{prop:finitespin}
\textbf{Finite spin exchange (Polynomial residues)}

At each weakly-coupled mass level $m_n$, our amplitudes exchange states with a finite set of spins less than or equal to some maximum spin $J_n \geq 0$. In other words, the residues of each amplitude at $s = m_n^2$ (with fixed $u < 0$) are polynomials in~$u$ with degree $J_n$. In the $stu$-symmetric case, only even spins are exchanged, so $J_n$ is even and the residue is a polynomial in $u^2$. If there are no massless poles, we set ${J_0 = -1}$ ($su$-symmetric) or ${J_0 = -2}$ ($stu$-symmetric) by convention so that our formulas are true in every case.

We believe that our first six assumptions are uncontroversial and that only this last assumption is worthy of further discussion. While finite spin exchange is a common assumption in the modern S-matrix bootstrap literature~\cite{Haring:2023zwu, Geiser:2022icl, Cheung:2022mkw, Cheung:2023adk, Cheung:2023uwn, Cheung:2024uhn, Cheung:2024obl, Berman:2024kdh}, its exclusion from the set of bootstrap assumptions is also common~\cite{Caron-Huot:2020cmc, Caron-Huot:2021rmr, Albert:2022oes, Albert:2023jtd, Albert:2023seb, Albert:2024yap, Berman:2023jys, Berman:2024wyt, Huang:2022mdb}. Finite spin exchange is often stated as a fundamental physical requirement related to locality since local particle-like excitations can be decomposed into a finite number of angular momentum modes, unlike generic extended objects~\cite{Cheung:2022mkw, Geiser:2022icl, Cheung:2023adk, Cheung:2023uwn, Cheung:2024obl, Cheung:2024uhn, Huang:2022mdb}. This position seems reasonable, so we take finite spin exchange as a mathematical assumption.

In~\autoref{sec:disc}, we discuss amplitudes known as infinite spin towers, first described in~\cite{Caron-Huot:2020cmc}, which violate the assumption of finite spin exchange while satisfying the other bootstrap assumptions.

\end{enumerate}

\subsection{Poles, residues, and couplings}

Before we proceed, let us further examine the poles and residues at each weakly-coupled mass level to set our conventions for various normalizations. The residue of each pole can be decomposed in spin, and the coefficients of this decomposition are related to the coupling constants of the theory.

For the massive exchanged states, we can decompose the residue at ${s=m_n^2}$ into a sum of spin-$j$ Gegenbauer polynomials with ${0 \leq j \leq J_n}$ (and $j$ even in the $stu$-symmetric case):
\begin{align}
\label{eq:resMassive}
    A(s,u)
    &
    \widesim{ s \to m_n^2 }
    - \frac{1}{ s-m_n^2 } \,
    m_n^{6-d}
    \sum_{j = 0}^{J_n}
    g_{n,j}^{2} \,
    \mcP_{j}^{(d)} \Big( 1 + \frac{2u}{m_n^2} \Big)
    \, ,
\end{align}
where $g_{n,j}$ is the dimensionless three-point coupling between two massless scalars and the state with mass $m_n$ and spin $j$. Without loss of generality, we assume $g_{n,j}$ is real so that we can write $g_{n,j}^2$ instead of $|g_{n,j}|^2$ throughout. Schematically, this interaction vertex is given by
\begin{align}
    \begin{tikzpicture}[baseline=-0.5ex]
    \node (0) at (0,0) [circle, fill=black, inner sep=0pt, minimum size=6pt] {}; 
    \node (1) at (0*360/3:1) {};
    \node (2) at (1*360/3:1) {};
    \node (3) at (2*360/3:1) {};
    \node (4) at (0*360/3:0.5) [label=below:{$m_n, j$}] {};
    \draw[thick, double] (0)--(1);
    \draw[thick] (0)--(2);
    \draw[thick] (0)--(3);
    \end{tikzpicture}
    &\propto
    i g_{n,j}
    \, ,
\end{align}
where the single lines denote massless scalars and the double lines denote the massive state (with its Lorentz indices suppressed). The contraction of two such vertices with the massive spin-$j$ propagator produces the expected $s$-channel pole at $m_n^2$:
\begin{align}
    \begin{tikzpicture}[baseline=-0.5ex]
    \node (0L) at (0,0) [circle, fill=black, inner sep=0pt, minimum size=6pt] {}; 
    \node (0R) at (1.5,0) [circle, fill=black, inner sep=0pt, minimum size=6pt] {}; 
    \node (1L) at ($(0L)+(1*360/3:1)$) {};
    \node (2L) at ($(0L)+(2*360/3:1)$) {};
    \node (1R) at ($(0R)+(1*360/3+180:1)$) {};
    \node (2R) at ($(0R)+(2*360/3+180:1)$) {};
    \node (label) at (.75,0) [label=below:{$m_n, j$}] {};
    \draw[thick,double] (0L)--(0R);
    \draw[thick]        (0L)--(1L);
    \draw[thick]        (0L)--(2L);
    \draw[thick]        (0R)--(1R);
    \draw[thick]        (0R)--(2R);
    \end{tikzpicture}
    &=
    - \frac{1}{ s-m_n^2 } \,
    m_n^{6-d} \,
    g_{n,j}^{2} \,
    \mcP_{j}^{(d)} \Big( 1 + \frac{2u}{m_n^2} \Big)
    \, .
\end{align}
The full pole at $m_n^2$ contributes the following delta function to $\Im A(s,u)$,
\begin{align}
    \Im A(s,u)
    &
    \widesim{ s \to m_n^2 }
    \pi \, \delta(s-m_n^2) \, m_n^{6-d}
    \sum_{j = 0}^{J_n}
    g_{n,j}^{2} \,
    \mcP_{j}^{(d)} \Big( 1 + \frac{2u}{m_n^2} \Big)
    \, ,
\end{align}
which contributes to the spectral density functions as
\begin{align}
\label{eq:rhoMassive}
    \rho_j(s)
    &
    \widesim{ s \to m_n^2 }
    \delta(s-m_n^2) \, m_n^{2} \,
    \sum_{j' = 0}^{J_n}
    g_{n,j'}^{2} \,
    \delta_{j,j'}^{\phantom{2}}
    \, .
\end{align}
Positivity requires each squared coupling to be non-negative $g_{n,j}^2 \geq 0$. Without loss of generality, we further assume that ${g_{n,J_n}^2 > 0}$ is strictly positive for each maximum spin~$J_n$. If we had $g_{n,J_n}^2 = 0$, then we would simply redefine $J_n$.

For the massless exchanged states, we must write the residues in a slightly different form since we cannot simply take the $m_n \to 0$ limit of~\eqref{eq:resMassive} while leaving the coupling constant dimensionless. Instead we opt for dimensionful coupling constants and write the simple pole at $s=0$ as
\begin{align}
\label{eq:resMassless}
    A(s,u)
    &
    \widesim{ s \to 0 }
    - \frac{1}{s}
    \sum_{j = 0}^{J_0}
    \lambda_{j}^{2} \, u^{j}
    \, ,
\end{align}
where $\lambda_{j}$ is the three-point coupling between two massless scalars and the spin-$j$ massless state. This coupling has mass dimension ${[\lambda_{j}^2] = 6-d-2j}$. Symmetrizing this massless $s$\nobreakdash-channel pole produces the full contribution to  the amplitude from massless exchanges:
\begin{align}
\label{eq:masslesspole}
    A^{(su)}_{\text{massless}}(s,u)
    &=
    - \sum_{j = 0}^{J_0}
    \lambda_{j}^2 \,
    \bigg(
      \frac{u^{j}}{s} 
    + \frac{s^{j}}{u}
    \bigg)
    \, ,
\no \\
    A^{(stu)}_{\text{massless}}(s,u)
    &=
    - \sum_{\substack{ j = 0 \\ \mathclap{\text{even}} \mathstrut }}^{J_0}
    \lambda_{j}^2 \, 
    \frac{1}{2}
    \bigg(
      \frac{t^{j}+u^{j}}{s}
    + \frac{u^{j}+s^{j}}{t}
    + \frac{s^{j}+t^{j}}{u}
    \bigg)
    \, .
\end{align}
These general expressions agree with tree-level Feynman diagram calculations for massless spin-$0$ or spin-$1$ exchanges (with $su$-symmetry) and massless spin-$0$ or spin-$2$ exchanges (with $stu$-symmetry):
\begin{align}
    A^{(su)}_{\text{spin-0}}(s,u)
    &=
    - \lambda_0^2 \, 
    \bigg( \frac{1}{s} + \frac{1}{u} \bigg)
    \, ,
    &
    A^{(stu)}_{\text{spin-0}}(s,u)
    &=
    - \lambda_0^2 \,
    \bigg( \frac{1}{s}  + \frac{1}{t}  + \frac{1}{u}  \bigg)
    \, ,
\no \\
    A^{(su)}_{\text{spin-1}}(s,u)
    &=
    - \lambda_1^2 \,
    \bigg( \frac{u}{s} + \frac{s}{u} \bigg)
    \, ,
    &
    A^{(stu)}_{\text{spin-2}}(s,u)
    &=
    + \lambda_2^2 \,
    \bigg( \frac{tu}{s} + \frac{us}{t} + \frac{st}{u} \bigg)
    \, .
\end{align}
In both the $su$- and $stu$-symmetric cases, non-zero $\lambda_0$ indicates the presence of a cubic interaction for the massless scalars. In the $su$-symmetric case, non-zero $\lambda_1$ indicates the presence of a gluon, meaning the symmetry group under which the massless scalars transform is gauged. In the $stu$-symmetric case, non-zero $\lambda_2$ indicates the presence of a graviton, meaning the theory is gravitational. This coupling is related to Newton's gravitational constant by ${\lambda_2^2 = 8 \pi G}$.

At the level of the amplitude, non-zero $\lambda_{j}$ appears to be consistent for any spin~$j$ even though interacting massless particles with spins higher than two are known to be physically inconsistent~\cite{Weinberg:1980kq}. Interacting higher-spin massless states are of course consistent in curved spacetime~\cite{Vasiliev:1990en, Vasiliev:2003ev}, but we are working in Minkowski space. Moreover, gravitons must be singlets under any symmetry group~\cite{Hillman:2024ouy} and should not appear as intermediate states in the $su$-symmetric amplitudes. Hence, we should really only consider ${J_0 \leq 1}$ and ${J_0 \leq 2}$ for the $su$-symmetric and $stu$-symmetric cases, respectively. However, to be as general as possible, we can ignore these constraints and consider any value of $J_0$.

Let us conclude this discussion by writing explicit expressions for the couplings at any mass level in terms of the generic amplitude $A(s,u)$. For the massless states, we first use a contour integral around ${s = 0}$ to compute the residue and then take $j$ derivatives with respect to~$u$ to isolate the coefficient of~$u^j$:
\begin{align}
\label{eq:extractCouplMassless}
    \lambda_j^2
    =
    \bigg(
    \frac{1}{q!}
    \frac{ \p^j }{ \p u^j }
    \ointctrclockwise\limits_{s=0}
    \frac{ \mathrm{d}s }{ 2\pi i } \,
    A(s,u)
    \bigg)\bigg|_{u=0}
    \, .
\end{align}
For the massive states, we first use a contour integral around ${s = m_n^2}$ to compute the residue and then use the orthogonality relation for the Gegenbauer polynomials to isolate the coefficient of the spin-$j$ term. The orthogonality relation is
\begin{align}
\label{eq:GegOrth}
    \frac{1}{2} \,
    \mcN^{(d)}_{\phantom{j}} n_{j_1}^{(d)}
    \int_{-1}^{1}
    \mathrm{d}x \,
    (1-x^2)^{\frac{d-4}{2}} \,
    \mcP_{j_1}^{(d)}(x) \,
    \mcP_{j_2}^{(d)}(x)
    =
    \delta_{j_1,j_2}
    \, ,
\end{align}
where $n_j^{(d)}$ is defined in~\eqref{eq:Gegn} and
\begin{align}
\label{eq:GegN}
    \mcN^{(d)}
    =
    \frac{ 1 }
         { (16\pi)^{\frac{d-2}{2}} \, \Gamma(\frac{d-2}{2}) }
    \, .
\end{align}
Thus,
\begin{align}
\label{eq:extractCouplMassive}
    g_{n,j}^{2}
    =
    \frac{ \mcN^{(d)}_{\phantom{j}} n_j^{(d)} }
         { 2 \pi i}
    \int\limits_{u=-m_n^2}^{u=0}
    \frac{ \mathrm{d}u }{ m_n^2 }
    \Big(
    \frac{ {-4u} (u+m_n^2) }{ m_n^4 }
    \Big)^{\frac{d-4}{2}}
    \mcP_{j}^{(d)} \Big( 1 + \frac{2u}{m_n^2} \Big)
    \ointctrclockwise\limits_{s=m_n^2}
    \frac{ \mathrm{d}s }{ m_n^2 }
    \frac{ A(s,u) }{ m_n^{4-d} }
\end{align}
for any $m_n > 0$ in the weakly-coupled spectrum. Technically, we have written the couplings as functionals of the amplitude $A(s,u)$.

\subsection{Summary of results}

From a naive, purely field theoretic perspective, it may appear that the masses, spins, and couplings of weakly-coupled states are totally unconstrained. However, these parameters are not free. The bootstrap assumptions of unitarity, causality, and locality imply an infinite number of constraints on the masses and spins of the weakly-coupled spectrum. In both the $su$-symmetric case and the $stu$-symmetric case, we can derive a set of ``Sequential Spin Constraints" (SSC) and ``Sequential Mass Constraints" (SMC).

\subsubsection{Sequential Spin Constraints}

The SSC are linear bounds which constrain the maximum spins $J_n$ at each mass level~$m_n$ in terms of the the degree $\alpha$ of the polynomial bound~\eqref{eq:bound} and the maximum spins at lower mass levels. The exact form of the SSC differs slightly between the $su$-symmetric and $stu$-symmetric cases.

\begin{thm}[$su$-SSC]
\label{thm:suSSC} For any $su$-symmetric amplitude which satisfies our bootstrap assumptions, the maximum spins $J_n$ at mass $m_n$ are constrained by
\begin{align}
    J_n
    \leq
    \max( \alpha-1, J_0, J_1, \dots, J_{n-1} ) + 1
\end{align}
for all $m_n > 0$ in the weakly-coupled spectrum. Equivalently, the mass of the lightest spin-$j$ state must be smaller than the mass of the lightest spin-$(j+1)$ state for all ${j \geq \max( \alpha, J_0+1)}$.
\end{thm}

\begin{thm}[$stu$-SSC]
\label{thm:stuSSC}
For any $stu$-symmetric amplitude which satisfies our bootstrap assumptions, the maximum spins $J_n$ at mass $m_n$ are constrained by
\begin{align}
    J_n
    \leq
    \max( 2 \lfloor \tfrac{\alpha-1}{2} \rfloor, J_0, J_1, \dots, J_{n-1} ) + 2
\end{align}
for all $m_n > 0$ in the weakly-coupled spectrum. Equivalently, the mass of the lightest spin-$j$ state must be smaller than the mass of the lightest spin-$(j+2)$ state for all even ${j \geq \max(2 \lceil \tfrac{\alpha}{2} \rceil, J_0+2)}$.
\end{thm}

\subsubsection{Sequential Mass Constraints}

The SMC are non-linear bounds which further constrain the value of the smallest mass at which a spin-$j$ particle appears. To cleanly state the SMC, we define $\mu_j$ as the first mass level at which a spin-$j$ state is exchanged and $\mfn_j$ as the index of this mass level so that ${m_{\mfn_j} = \mu_j}$. Equivalently, $\mfn_j$ is the smallest value of $n$ such that $J_n \geq j$. The SMC bound ratios of $\mu_j$. The SMC inequality is identical for both the $su$-symmetric and $stu$-symmetric cases, but the conditions of applicability slightly differ.

\begin{thm}[$su$-SMC]
\label{thm:suSMC}
For any $su$-symmetric amplitude which satisfies our bootstrap assumptions, the masses $\mu_j$ of the lightest spin-$j$ particles are constrained by
\begin{align}
    \bigg(
    \frac{ \mu_{j''} }{ \mu_{j\phantom{''}} }
    \bigg)^{j'-j}
    \leq
    \bigg(
    \frac{ \mu_{j'} }{ \mu_{j\phantom{'}} }
    \bigg)^{j''-j}
\end{align}
for all ${j'' \geq j' \geq j \geq \max(\alpha, J_0+1)}$ with ${\mu_j,\mu_{j'},\mu_{j''} > 0}$ in the weakly-coupled spectrum.
\end{thm}

\begin{thm}[$stu$-SMC]
\label{thm:stuSMC}
For any $stu$-symmetric amplitude which satisfies our bootstrap assumptions, the masses $\mu_j$ of the lightest spin-$j$ particles are constrained by
\begin{align}
    \bigg(
    \frac{ \mu_{j''} }{ \mu_{j\phantom{''}} }
    \bigg)^{j'-j}
    \leq
    \bigg(
    \frac{ \mu_{j'} }{ \mu_{j\phantom{'}} }
    \bigg)^{j''-j}
\end{align}
for all even ${j'' \geq j' \geq j \geq \max(2 \lceil \tfrac{\alpha}{2} \rceil, J_0+2)}$ with ${\mu_j,\mu_{j'},\mu_{j''} > 0}$ in the weakly-coupled spectrum.
\end{thm}

\subsection{Outline}

The remainder of this paper is organized as follows.
\begin{itemize}
    \item In \autoref{sec:su}, we prove the SSC and SMC for $su$-symmetric amplitudes.
    \item In \autoref{sec:stu}, we prove the SSC and SMC for $stu$-symmetric amplitudes.
    \item In \autoref{sec:SYMSUGRA}, we discuss the extension of our results to super-gluon and super-graviton amplitudes, such as the amplitudes of open and closed superstrings.
    \item Finally, we conclude in \autoref{sec:disc}.
\end{itemize}
The proofs in \autoref{sec:su} and \autoref{sec:stu} are written in parallel so that each section can be read independently of the other.

\section{The \texorpdfstring{$su$}{su}-symmetric case}
\label{sec:su}

In this section, we prove the Sequential Spin Constraints (SSC) and the Sequential Mass Constraints (SMC) for all $su$-symmetric amplitudes that satisfy our bootstrap assumptions. We begin by defining a pole-subtracted amplitude whose first mass level is at $m_n$ so that we can isolate the contribution of each state. We then carefully analyze this amplitude's Wilson coefficients. Our derivation of the dispersion relations for these Wilson coefficients is not new~\cite{Arkani-Hamed:2020blm, Bellazzini:2020cot}, but the constraints on the spectrum (i.e.\ the SSC and SMC) are new to the literature as discussed in \autoref{sec:intro}.

\subsection{The pole-subtracted amplitude}

For any mass $m_n > 0$ in the weakly-coupled spectrum of the $su$-symmetric amplitude $A^{(su)}(s,u)$, we define the pole-subtracted amplitude $A^{(su)}_n(s,u)$ by explicitly subtracting the massless poles and the massive poles with masses $m_p < m_n$:
\begin{align}
\label{eq:polesub-su}
    A^{(su)}_n(s,u)
    &=
    A^{(su)}(s,u)
    +
    \sum_{j = 0}^{J_0}
    \lambda_{j}^2
    \bigg(
      \frac{u^{j}}{s} 
    + \frac{s^{j}}{u}
    \bigg)
\\ \no
    & \quad
    +
    \sum_{p = 1}^{n-1}
    m_p^{6-d}
    \sum_{j = 0}^{J_p}
    g_{p,j}^{2}
    \bigg[
    \frac{1}{ s-m_p^2 } \,
    \mcP_{j}^{(d)} \Big( 1 + \frac{2u}{m_p^2} \Big)
    +
    \frac{1}{ u-m_p^2 } \,
    \mcP_{j}^{(d)} \Big( 1 + \frac{2s}{m_p^2} \Big)
    \bigg]
    \, .
\end{align}
The ``subtractions" are accomplished with plus signs because the poles have an overall minus sign as described in~\eqref{eq:resMassive} and~\eqref{eq:resMassless}. We also recall that $J_p$ is the maximum spin exchanged at mass~$m_p$.

This pole-subtracted amplitude is a well-defined functional of the original amplitude since the couplings $\lambda_j^2$ and $g_{p,j}^2$ can be written in terms of $A^{(su)}(s,u)$ using~\eqref{eq:extractCouplMassless} and~\eqref{eq:extractCouplMassive}. The subtraction manifestly respects the $su$-symmetry of the original amplitude while eliminating the simple poles in $s$ at $s=0,m_1^2, \dots, m_{n-1}^2$. Otherwise, the pole-subtracted amplitude has the same analytic structure as the original amplitude, depicted in~\autoref{fig:analyticity}. 

\subsubsection{Polynomial bound}

Due to the various powers of~$s$ in the numerators of the $u$-channel poles, the pole-subtracted amplitude is polynomially-bounded by
\begin{align}
    \lim_{\substack{ |s| \to \infty \\
                     \mathclap{\text{fixed } u<0 }}}
    \frac{A^{(su)}_n(s,u)}
         {s^{\mcJ_n}}
    = 0
\end{align}
with
\begin{align}
\label{eq:Jn-su}
    \mcJ_n 
    =
    \max(\alpha-1, J_0, J_1, \dots, J_{n-1})+1
    \, ,
\end{align}
where $\alpha$ is the degree of the polynomial bound of the original amplitude $A^{(su)}(s,u)$ in~\eqref{eq:bound}. This polynomial bound $\mcJ_n$ is finite since each maximum spin $J_n$ is finite (by assumption). If there are no massless poles, we set $J_0 = -1$ by convention.

\subsubsection{Low-energy expansion}

Since we have subtracted the massless poles, the low-energy expansion near $s=u=0$ is analytic in $s$ and $u$ and given by
\begin{align}
\label{eq:lowenergy-su}
    A^{(su)}_n(s,u) 
    &=
    \sum_{k = 0}^\infty
    \sum_{q = 0}^k
    a^{(su)}_{n;k,q} \,
    s^{k-q} u^q
    \, .
\end{align}
This low-energy expansion is valid for $|s|, |u| < m_n^2$. As a consequence of $su$-symmetry, the Wilson coefficients obey homogeneous linear equations, graded by the Mandelstam order~$k$. Namely, $a^{(su)}_{n;k,q} = a^{(su)}_{n;k,k-q}$ for all $k \geq q \geq 0$.

\subsubsection{Dispersion relations}

\begin{figure}
\centering
\begin{tikzpicture}
    \clip (-5.5,-5.5) rectangle (5.5,5.5);
    
    \draw[fill=gray!20, color=gray!20]
         (-5.2,-5.4) --
         (-5.2, 5.4) --
         ( 5.2, 5.4) --
         ( 5.2,-5.4) -- cycle;
    
    \coordinate (m0) at ( 0.0, 0.0);
    \coordinate (m1) at ( 1.4, 0.0);
    \coordinate (mn) at ( 2.7, 0.0);
    \coordinate (mL) at ( 4.0, 0.0);

    \coordinate (u)  at (-0.6, 0.0);
    \coordinate (u0) at ($-1*(u)-(m0)$);

    \node at ( 4.5, 0.5) {$\scriptstyle\Re(s)$};
    \node at (-0.6, 4.6) {$\scriptstyle\Im(s)$};

    \coordinate (xr) at ( 0.3, 0.0);
    \coordinate (yr) at ( 0.0, 0.2);
    
    \draw     (-5.0, 0.0)   -- ($(mn)+(xr)$);
    \draw[->] ($(mL)-(xr)$) -- ( 5.0, 0.0);

    \draw     ( 0.0,-5.0)   -- ( 0.0,-0.75);
    \draw[->] ( 0.0, 0.0)   -- ( 0.0, 5.0);

    \draw (m0) -- ($(m0)-(yr)$);
    \draw (mn) -- ($(mn)-(yr)$);
    \draw (mL) -- ($(mL)-(yr)$);
    \draw (u0) -- ($(u0)-(yr)$);
    
    \node at ($(m0)-2.5*(yr)$) {$\scriptstyle 0 \mathstrut$};
    \node at ($(mn)-2.5*(yr)$) {$\scriptstyle m_n^2$};
    \node at ($(mL)-2.5*(yr)$) {$\scriptstyle \Lambda_{\phantom{n}}^2$};

    \node at ($(u0)-2.5*(yr)$) {$\scriptstyle -u \phantom{-}$};

    \node at ($0.5*(mL)+0.5*(mn)$) {$\cdots$};

    \filldraw[color=red] (m0) circle (2pt);
    \filldraw[color=red] (mn) circle (2pt);
    \filldraw[color=red] (mL) circle (1pt);

    \draw[color=red, very thick, decorate,
          decoration={zigzag, amplitude=2pt, segment length=5pt}]
        (mL) -- ( 4.95, 0.0);

    \draw[color=blue, thick, postaction={decorate},
          decoration={markings, mark=at position 0.125 with {\arrow{>}}}]
         (m0) circle [radius=7pt];
    \draw[color=blue, thick, postaction={decorate},
          decoration={markings, mark=at position 0.125 with {\arrow{>}}}]
         (mn) circle [radius=7pt];

    \draw[color=blue, thick] ($(mL)+(0,7pt)$) arc ( 90:270:7pt);

    \draw[color=blue, thick, postaction={decorate},
          decoration={markings, mark=at position 0.25 with {\arrow{<}}}]
         ($(mL)+(0,7pt)$) -- ($(5,0)+(0,7pt)$);
    \draw[color=blue, thick, postaction={decorate},
          decoration={markings, mark=at position 0.25 with {\arrow{>}}}]
         ($(mL)-(0,7pt)$) -- ($(5,0)-(0,7pt)$);

    \draw[color=blue, thick] ($(-5,0)-(0,7pt)$) -- ($(-5,0)+(0,7pt)$);

    \draw[color=blue, thick, postaction={decorate},
          decoration={markings, mark=at position 0.2 with {\arrow{<}},
                                mark=at position 0.8 with {\arrow{<}}}]
         ($(5,0)+(0,7pt)$) arc (0:180:5);
    \draw[color=blue, thick, postaction={decorate},
          decoration={markings, mark=at position 0.25 with {\arrow{<}},
                                mark=at position 0.75 with {\arrow{<}}}]
         ($(-5,0)-(0,7pt)$) arc (-180:0:5);

    \node[fill=white] at ( 3.8, 4.5)
        {$\displaystyle\frac{A^{(su)}_n(s,u)}{s^{k-q+1}}$};
\end{tikzpicture}

\caption{The contour integral in~\eqref{eq:contour-su}.}

\label{fig:contour-su}

\end{figure}
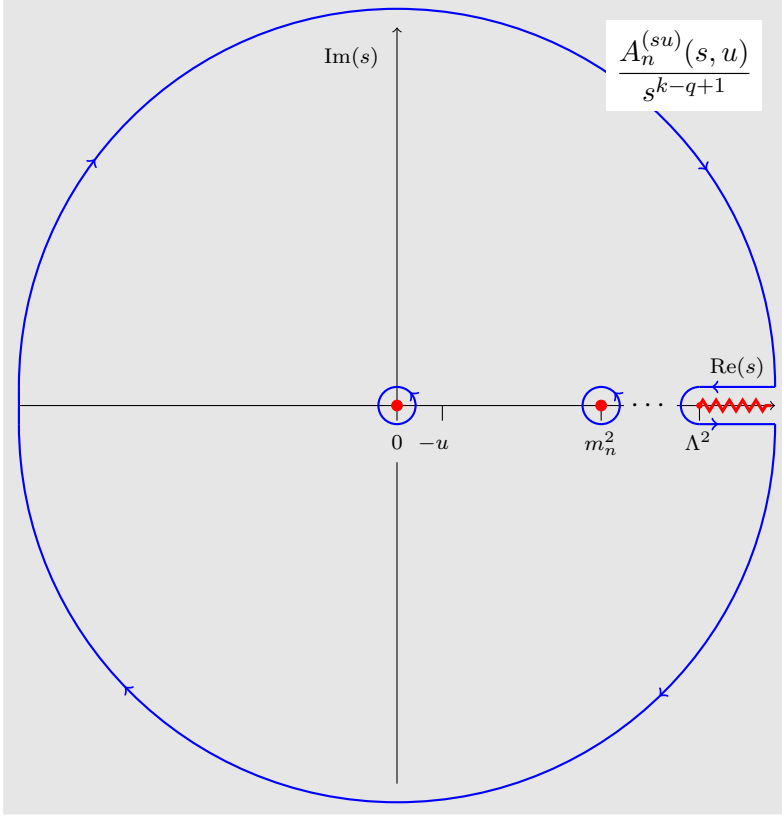

We can derive a dispersion relation for the the Wilson coefficient $a^{(su)}_{n;k,q}$ with $k-q \geq \mcJ_n$ by considering the following contour integral at fixed $u \in (-m_n^2, 0)$:
\begin{align}
\label{eq:contour-su}
    \Bigg\{
    \quad
    \ointctrclockwise\limits_{\mathclap{s=0^{\phantom{2}}}}
    \quad \,
    +
    \sum_{m_p \geq m_n}
    \quad \,
    \ointctrclockwise\limits_{\mathclap{s=m_p^2}}
    \quad \,
    +
    \quad \,
    \ointctrclockwise\limits_{\mathclap{s \in [\Lambda^2,\infty] }}
    \quad \,
    +
    \quad \,
    \varointclockwise\limits_{\mathclap{s=\infty^{\phantom{2}}}}
    \quad
    \Bigg\}
    \frac{ \mathrm{d}s }{ 2\pi i }
    \frac{ A^{(su)}_n(s,u) }{ s^{k-q+1} }
    = 0
    \, .
\end{align}
The full contour is depicted in~\autoref{fig:contour-su} and trivially vanishes by Cauchy's integral theorem.
\begin{itemize}
\item The contour around $s = 0$ extracts the Wilson coefficients in~\eqref{eq:lowenergy-su}:
\begin{align}
\label{eq:contour-su1}
    \ointctrclockwise\limits_{\mathclap{s=0}}
    \,
    \frac{ \mathrm{d}s }{ 2\pi i }
    \frac{ A^{(su)}_n(s,u) }{ s^{k-q+1} }
    &=
    \sum_{q' = 0}^\infty
    a^{(su)}_{n;k-q+q',q'} \,
    u^{q'}
    \, .
\intertext{\item The contours around each $s=m_p^2$ can be evaluated using~\eqref{eq:resMassive}:}
\label{eq:contour-su2}
    \ointctrclockwise\limits_{\mathclap{s=m_p^2}}
    \,
    \frac{ \mathrm{d}s }{ 2\pi i }
    \frac{ A^{(su)}_n(s,u) }{ s^{k-q+1} }
    &=
    -
    \frac{1}{m_p^{2k-2q+d-4}}
    \sum_{ j = 0 }^{J_p}
    g_{p,j}^{2} \,
    \mcP_{j}^{(d)} \Big( 1 + \frac{2u}{m_p^2} \Big)
    \, .
\intertext{\item The contour around $s \in [\Lambda^2, \infty]$ computes the discontinuity of the integrand across the real $s$-axis, and within this integral, the imaginary part of the pole-subtracted amplitude is equal to the imaginary part of the original amplitude $A^{(su)}(s,u)$, which can be expanded as a sum over the spectral density functions $\rho_j(s)$ defined in~\eqref{eq:rho}:}
\label{eq:contour-su3}
    \ointctrclockwise\limits_{\mathclap{s \in [\Lambda^2,\infty] }}
    \,
    \frac{ \mathrm{d}s }{ 2\pi i }
    \frac{ A^{(su)}_n(s,u) }{ s^{k-q+1} }
    &=
    -
    \int_{\Lambda^2}^\infty
    \frac{ \mathrm{d}s }{ s^{k-q+1} }
    \sum_{j = 0}^\infty
    \rho_j(s) \,
    \mcP_j^{(d)} \Big( 1 + \frac{2u}{s} \Big)
    \vphantom{\frac{ \Im A^{(su)}(s,u) }{ s^{k-q+1} }}
    \, .
\end{align}
\item Finally, the contour at infinity vanishes since $k-q \geq \mcJ_n$.
\end{itemize}
Each contribution to the full contour integral is a power series in~$u$, valid for all values of $u \in (-m_n^2, 0)$ with no obstruction to taking the limit ${u \to 0}$. Thus, we can act with the operator $\frac{1}{q!} (\partial / \partial u)^q$ on the full contour integral before taking ${u \to 0}$ to extract the coefficient of $u^q$.

After some straightforward algebra and a change of integration variables to the dimensionless combination ${z = \Lambda^2 s^{-1}}$, we arrive at the following dispersion relation for the Wilson coefficients of $A_n^{(su)}(s,u)$. This relation holds for all integers $n \geq 1$ with $m_n$ in the weakly-coupled spectrum and for all integers $k,q \geq 0$ satisfying $k-q \geq \mcJ_n$:
\begin{align}
\label{eq:disprel-su}
    a^{(su)}_{n;k,q}
    &=
    \sum_{m_p \geq m_n}
    \frac{1}{m_p^{2k+d-4}}
    \sum_{ j = q }^{J_p}
    g_{p,j}^{2} \,
    v_{j,q}^{(d)}
\no \\
    & \quad
    +
    \frac{1}{\Lambda^{2k+d-4}}
    \int_0^1
    \mathrm{d}z \,
    z^{k-1+\frac{d-4}{2}}
    \sum_{j=q}^\infty
    \rho_j \big( \Lambda^2 z^{-1} \big) \,
    v_{j,q}^{(d)}
    \, .
\end{align}
Here $v_{j,q}^{(d)}$ is the coefficient of $x^q$ in the spin-$j$ Gegenbauer polynomial $\mcP_j^{(d)}(1+2x)$:
\begin{align}
\label{eq:Gegv}
    v_{j,q}^{(d)}
    &=
    \binom{j}{q}
    \frac{ \Gamma(j+d-3+q) \, \Gamma(\frac{d-2}{2}) }
         { \Gamma(j+d-3) \, \Gamma(\frac{d-2}{2}+q) }
    \, .
\end{align}
These coefficients are manifestly non-negative for all $d > 3$ and all integers $j,q \geq 0$.

We could have alternatively computed this dispersion relation using a contour which encircles $s \in [m_n^2,\infty]$ (without closing the contours around the poles at each $s=m_p^2$). In this case, we find
\begin{align}
\label{eq:disprel-su-alt}
    a^{(su)}_{n;k,q}
    &=
    \frac{1}{m_n^{2k+d-4}}
    \int_0^1
    \mathrm{d}z_n \,
    z_n^{k-1+\frac{d-4}{2}}
    \sum_{j=q}^\infty
    \rho_j \big( m_n^2 z_n^{-1} \big) \,
    v_{j,q}^{(d)}
    \, ,
\end{align}
where the new integration variable is defined as ${z_n = m_n^2 s^{-1}}$. For $s \in [m_n^2, \Lambda^2)$, the spectral density functions are given by a sum over delta functions with support at each ${s=m_p^2}$, as in~\eqref{eq:rhoMassive}, which reproduces the previous form of the dispersion relation.

The sums over $j$ in~\eqref{eq:disprel-su} and~\eqref{eq:disprel-su-alt} begin at $j=q$ because the coefficients $\smash{v_{j,q}^{(d)}}$ vanish for integers ${0 \leq j < q}$. Hence, only spin-$q$ states and higher contribute to $a^{(su)}_{n;k,q}$. We also note that these Wilson coefficients are manifestly non-negative as a consequence of the positivity of the spectral densities ${\rho_j(s) \geq 0}$ and the positivity of each $v_{j,q}^{(d)}  \geq 0$.

\subsubsection{The \texorpdfstring{$k \to \infty$}{k to infinity} limit}

Before we proceed, let us examine the $k \to \infty$ limit (with $q$ fixed) of the Wilson coefficients using the dispersion relation~\eqref{eq:disprel-su}. We can safely take this limit because each contribution to~\eqref{eq:disprel-su} is non-negative. However, we must first eliminate the overall mass dimension. Given any mass $m$, we can form the dimensionless combination $m^{2k+d-4} \, a^{(su)}_{n;k,q}$. The choice of this mass determines whether the limit is zero, non-zero and finite, or infinite.
\begin{itemize}
\item If $m < m_{n}$, then the limit vanishes:
\begin{align}
\label{eq:klim-su-1}
    \lim_{k \to \infty}
    \Big(
    m^{2k+d-4} \, a^{(su)}_{n;k,q}
    \Big)
    &=
    0
    \, .
\end{align}
\item If $m = m_{n}$, then the limit results in a finite sum over the spin-$j$ couplings at that mass level beginning with $j=q$,
\begin{align}
\label{eq:klim-su-2}
    \lim_{k \to \infty}
    \Big(
    m_{n}^{2k+d-4} \, a^{(su)}_{n;k,q}
    \Big)
    &=
    \sum_{ j = q }^{J_n}
    g_{n,j}^{2} \, 
    v_{j,q}^{(d)}
    \, ,
\end{align}
which vanishes if $q > J_n$. This sum is finite because (by assumption) the amplitude exchanges a finite set of spins with maximum spin $J_n$ at mass $m_n$.
\item If $m > m_{n}$, then the limit generically diverges, but if $m = \mu_q$ is the first mass level at which a spin-$q$ or higher state is exchanged (with $\mu_q > m_n$, assuming that such a mass level exists), then the limit results in a finite sum over the spin-$j$ couplings at that mass level beginning with $j=q$, similar to the previous equation:
\begin{align}
\label{eq:klim-su-3}
    \lim_{k \to \infty}
    \Big(
    \mu_q^{2k+d-4} \, a^{(su)}_{n;k,q}
    \Big)
    &=
    \sum_{ j = q }^{J_{\mfn_q}}
    g_{\mfn_q,j}^{2} \, 
    v_{j,q}^{(d)}
    \, ,
\end{align}
where $\mfn_q$ is the index of the first mass level at which a spin-$q$ state is exchanged so that ${m_{\mfn_q} = \mu_q}$. This limit is finite because, in this case, no terms proportional to $x^k$ with ${x>1}$ appear after multiplying the dispersion relation~\eqref{eq:disprel-su} by $\mu_q^{2k+d-4}$.
\end{itemize}

\subsubsection{An inequality from positivity}
\label{sec:jensen}

Our next step will be to use the positivity of the spectral densities and the alternative form of the dispersion relation~\eqref{eq:disprel-su-alt} to write an inequality relating Wilson coefficients with different values of the ``$k$" subscript. Let us first consider any sequence of real numbers $(A_0, A_1, A_2, \dots)$ with $A_0 = 1$ and
\begin{align}
    A_i
    =
    \int_0^1
    \mathrm{d}z \,
    \sigma(z) \, z^i
    \, ,
\end{align}
where $\sigma(z)$ is a positive measure satisfying the distributional inequality ${0 \leq \sigma(z) \leq 1}$ so that $A_{i+1} \leq A_i$ with each $A_i \in [0, 1]$. (The upper bound on $\sigma(z)$ is a consequence of the normalization convention $A_0=1$ and is unrelated to the positivity ${\sigma(z) \geq 0}$.) Such a sequence is known as a Hausdorff moment sequence in the mathematics literature, but we will not use any special mathematical results here. For any integer $q \geq 0$ and any mass $m$, the following ratios of Wilson coefficients form a Hausdorff moment sequence:
\begin{align}
\label{eq:Ai-su}
    A_i
    =
    m^{2i} \,
    \frac{ a^{(su)}_{n;q+\mcJ_n+i,q} }
         { a^{(su)}_{n;q+\mcJ_n,q \phantom{+i}} }
    \, .
\end{align}
In this case, positivity of the measure $\sigma(z)$ follows from the alternative form of the dispersion relation~\eqref{eq:disprel-su-alt} and the positivity of the spectral density functions.

Now consider any two integers $i_2 > i_1 \geq 1$. The function $z \mapsto z^{i_2/i_1}$ is convex on the domain~$[0,1]$. Thus, by Jensen's integral inequality (a generalization of the Cauchy–Schwarz integral inequality which relates the value of a convex function of an integral to the integral of the convex function)~\cite{10.5555/26851}, we have
\begin{align}
    A_{i_1}^{i_2}
    =
    \bigg(
    \int_0^1
    \mathrm{d}z \,
    \sigma(z) \, z^{i_1}
    \bigg)^{i_2/i_1 \cdot \, i_1}
    \leq 
    \bigg(
    \int_0^1
    \mathrm{d}z \,
    \sigma(z) \,
    ( z^{i_1} )^{i_2/i_1}
    \bigg)^{i_1}
    =
    A_{i_2}^{i_1}
\end{align}
for any Hausdorff moment sequence $(A_0, A_1, A_2, \dots)$. This inequality trivially extends to all integers ${i_2 \geq i_1 \geq 0}$. If we now substitute into this inequality the sequence of ratios of Wilson coefficients~\eqref{eq:Ai-su} and choose $q=k-\mcJ_n-i_1$ for any integer $k \geq 2\mcJ_n+i_1$, we find
\begin{align}
    \big(
    a^{(su)}_{n;k,k-\mcJ_n-i_1}
    \big)^{i_2}
    &
    \leq
    \big(
    a^{(su)}_{n;k+i_2-i_1,k-\mcJ_n-i_1}
    \big)^{i_1}
    \big(
    a^{(su)}_{n;k-i_1,k-\mcJ_n-i_1}
    \big)^{i_2-i_1}
    \, .
\end{align}
This inequality does not depend on the arbitrary mass $m$ which appeared in~\eqref{eq:Ai-su}, but the variable $k$ appears in both the ``$k$" subscript and the ``$q$" subscript of each Wilson coefficient. To simplify the dependence on $k$, we can use crossing symmetry. Recall that $su$-symmetry implies that the Wilson coefficients obey $a^{(su)}_{n;k,q} = a^{(su)}_{n;k,k-q}$ for all ${k \geq q \geq 0}$. Using this relation in the previous inequality yields
\begin{align}
\label{eq:ineq-su}
    \big(
    a^{(su)}_{n;k,\mcJ_n+i_1}
    \big)^{i_2}
    &
    \leq
    \big(
    a^{(su)}_{n;k+i_2-i_1,\mcJ_n+i_2}
    \big)^{i_1}
    \big(
    a^{(su)}_{n;k-i_1,\mcJ_n}
    \big)^{i_2-i_1}
    \, .
\end{align}
Now $k$ only appears in the ``$k$" subscript of each Wilson coefficient. Moreover, each of these Wilson coefficients has a dispersion relation of the form~\eqref{eq:disprel-su} for all integers ${i_2 \geq i_1 \geq 0}$ because $k \geq 2\mcJ_n+i_1$ implies that the difference between the ``$k$" subscript and the ``$q$" subscript is greater than or equal to $\mcJ_n$ for all three coefficients.

\subsection{Proving the \texorpdfstring{$su$}{su}-SSC and \texorpdfstring{$su$}{su}-SMC}

Both the $su$-symmetric SSC and SMC follow directly from the inequality~\eqref{eq:ineq-su}. For each proof, we simply multiply both sides of the inequality by an appropriate factor and then analyze the $k \to \infty$ limit.

\subsubsection{Proving the \texorpdfstring{$su$}{su}-SSC}

The $su$-SSC can be written as $J_n \leq \mcJ_n$. In other words, the maximum spin $J_n$ exchanged at mass $m_n$ is less than or equal to $\mcJ_n$, the degree of the polynomial bound in~\eqref{eq:Jn-su}. 

We shall prove this statement by contradiction. Let us first assume that $J_n \geq \mcJ_n + 1$. Next we consider the inequality~\eqref{eq:ineq-su} and multiply both sides by $m_n^{(2k+d-4)i_2}$. We then set ${i_1 = J_n - \mcJ_n}$ and ${i_2 = i_1 + 1}$ so that ${i_2 > i_1 \geq 1}$. We can now compute the $k \to \infty$ limit using~\eqref{eq:klim-su-2} to find
\begin{align}
    \Bigg(
    \sum_{ j = J_n }^{ J_n }
    g_{n,j}^{2} \, 
    v_{j,J_n}^{(d)}
    \Bigg)^{J_n-\mcJ_n+1}
    &\leq
    \Bigg(
    \sum_{ j = J_n+1 }^{ J_n }
    g_{n,j}^{2} \, 
    v_{j,J_n+1}^{(d)}
    \Bigg)^{J_n-\mcJ_n}
    \Bigg(
    \sum_{ j = \mcJ_n }^{ J_n }
    g_{n,j}^{2} \, 
    v_{j,\mcJ_n}^{(d)}
    \Bigg)
    \, .
\end{align}
The first sum on the right-hand side vanishes identically because the lower limit is larger than the upper limit, and the sum on the left-hand side is manifestly non-negative. Since the coefficient $v_{j,j}^{(d)} > 0$, we must have $g_{n,J_n}^2 = 0$, which contradicts our bootstrap assumptions. Recall that in~\autoref{sec:intro} we assumed that the coupling of the maximum spin at each mass level $g_{n,J_n}^2 > 0$ was strictly positive without loss of generality. Having reached a contradiction, we conclude that $J_n \leq \mcJ_n$.

\subsubsection{Proving the \texorpdfstring{$su$}{su}-SMC}

To prove the $su$-SMC, we begin with a corollary of the $su$-SSC. If $\mfn_j$ is the index of the first mass level at which a spin-$j$ state is exchanged (so that $m_{\mfn_j} = \mu_j$), then the $su$-SSC implies that the maximum spin at this mass level is equal to $j$. More precisely, ${J_{\mfn_j} = j}$ and ${\mcJ_{\mfn_j} = j}$ for all ${j \geq \max(\alpha,J_0+1)}$ with $\mu_j$ in the weakly-coupled spectrum. In other words, the first appearance of a spin-$j$ state must be at a mass level at which it is the maximum spin.

This corollary then implies that the limit~\eqref{eq:klim-su-3} with ${q = j}$ for any ${j \geq \max(\alpha,J_0+1)}$ such that $\mu_j \geq m_n$ simplifies to
\begin{align}
\label{eq:klim-su-4}
    \lim_{k \to \infty}
    \Big(
    \mu_j^{2k+d-4} \, a^{(su)}_{n;k,j}
    \Big)
    &=
    g_{\mfn_j,j}^{2} \, 
    v_{j,j}^{(d)}
    > 0
    \, ,
\end{align}
which is strictly positive because both $v_{j,j}^{(d)} > 0$ and $g_{\mfn_j,j}^{2} = g_{\mfn_j,J_{\mfn_j}}^2 > 0$ are positive.

Now we consider the inequality~\eqref{eq:ineq-su}, setting ${n = \mfn_j}$ for ${j \geq \max(\alpha,J_0+1)}$ with $\mu_j$ in the weakly-coupled spectrum so that $\mcJ_n = j$. We define ${j' = j+i_1}$, and ${j'' = j+i_2}$ with $\mu_{j'}, \mu_{j''}$ also in the weakly-coupled spectrum. After some rearranging,~\eqref{eq:ineq-su} becomes
\begin{align}
    1 \leq
    \frac{
    \big(
    a^{(su)}_{\mfn_j;k+j''-j',j''}
    \big)^{j'-j}
    \big(
    a^{(su)}_{\mfn_j;k+j-j',j}
    \big)^{j''-j'}
    }
    {
    \big(
    a^{(su)}_{\mfn_j;k,j'}
    \big)^{j''-j}
    }
    \, .
\end{align}
Next we multiply both sides of this expression by
\begin{align}
    \frac{
    \big(
    \mu_{j''}^{2(k+j''-j')+d-4}
    \big)^{j'-j}
    \big(
    \mu_{j \vphantom{j'}}^{2(k+j-j')+d-4}
    \big)^{j''-j'}
    }
    {
    \big(
    \mu_{j'}^{2k+d-4}
    \big)^{j''-j}
    }
\end{align}
to find a messy inequality,
\begin{align}
\label{eq:su-SMC-ineq}
    & \quad
    \bigg(
    \frac{ \mu_{j''}^2 }{ \mu_{j\phantom{''}}^2 }
    \bigg)^{(j''-j')(j'-j)}
    \Bigg(
    \bigg(
    \frac{ \mu_{j''}^2 }{ \mu_{j\phantom{''}}^2 }
    \bigg)^{j'-j}
    \bigg(
    \frac{ \mu_{j\phantom{'}}^2 }{ \mu_{j'}^2 }
    \bigg)^{j''-j}
    \Bigg)^{k + \frac{d-4}{2} }
\no \\[1.5ex]
    & \qquad \qquad
    \leq
    \frac{
    \big(
    \mu_{j''}^{2(k+j''-j')+d-4} \,
    a^{(su)}_{\mfn_j;k+j''-j',j''}
    \big)^{j'-j}
    \big(
    \mu_{j \vphantom{j'}}^{2(k+j-j')+d-4} \,
    a^{(su)}_{\mfn_j;k+j-j',j}
    \big)^{j''-j'}
    }
    {
    \big(
    \mu_{j'}^{2k+d-4} \,
    a^{(su)}_{\mfn_j;k,j'}
    \big)^{j''-j}
    }
    \, .
\end{align}
The left-hand side is positive for all finite $k$. The $k \to \infty$ limit of the right-hand side can be computed using~\eqref{eq:klim-su-4}. The result is finite and positive. Thus, the same limit of the left-hand side must also be finite. Hence, the quantity inside the large parentheses in~\eqref{eq:su-SMC-ineq} must be less than or equal to one. Therefore,
\begin{align}
\label{eq:su-SMC}
    \bigg(
    \frac{ \mu_{j''}^2 }{ \mu_{j\phantom{''}}^2 }
    \bigg)^{j'-j}
    \leq
    \bigg(
    \frac{ \mu_{j'}^2 }{ \mu_{j\phantom{'}}^2 }
    \bigg)^{j''-j}
    \, .
\end{align}
This completes our proof of the $su$-symmetric SMC.

\subsection{Saturating the \texorpdfstring{$su$}{su}-SSC and \texorpdfstring{$su$}{su}-SMC}

We conclude this section by discussing the saturation of the $su$-SSC and $su$-SMC bounds.

\subsubsection{Saturating the \texorpdfstring{$su$}{su}-SSC}

The $su$-SSC is a simple linear inequality, and it is trivially saturated by $J_n = \mcJ_n$ for all $n \geq 1$ with $m_n$ in the weakly-coupled spectrum. In this case, the maximum spins at each mass level increase linearly as $J_n = n+\hat{\jmath}-1$, where $\hat{\jmath} = \max(\alpha,J_0+1)$, so that the masses obey ${m_n = \mu_{n+\hat{\jmath}-1}}$. In summary, when the $su$-SSC is saturated, the maximum spins increase by one at each subsequent mass level beginning with spin-$\hat{\jmath}$ at mass $m_1$.

The open superstring spectrum in \autoref{fig:regge} saturates the $su$-SSC with ${J_0 = \alpha = 1}$ and ${J_n = n+1}$. We revisit this observation in \autoref{sec:SYMSUGRA}.

\subsubsection{Saturating the \texorpdfstring{$su$}{su}-SMC}

The $su$-SMC is more complicated because~\eqref{eq:su-SMC} is highly non-linear. Moreover, the $su$-SMC only constrains the set of masses $\mu_j$ at which a spin-$j$ state is first exchanged. In general, this set is a subset of the full set of masses $m_n$. That is,
\begin{align}
\label{eq:subset-su}
    \{ \mu_j : j \geq \hat{\jmath} \}
    \subset
    \{ m_n : n \geq 1 \}
    \, .
\end{align}
The smallest spin for which the $su$-SMC applies is ${\hat{\jmath} = \max(\alpha,J_0+1)}$, and for all $j \geq \hat{\jmath}$, the following masses saturate~\eqref{eq:su-SMC}:
\begin{align}
\label{eq:su-SMC-sol}
    \mu_j^2
    =
    \mu_{ \hat{\jmath} }^2 \,
    \bigg(
    \frac{ \mu_{ \hat{\jmath}+1 }^2 }
         { \mu_{ \hat{\jmath} \phantom{+1} }^2 }
    \bigg)^{j-\hat{\jmath}}
    \, .
\end{align}
Since we are free to measure masses in units of $\mu_{ \hat{\jmath} }$, this expression has one free parameter, the ratio of masses $\chi = \mu_{ \hat{\jmath}+1 }^2 / \mu_{ \hat{\jmath}}^2$. As a consequence of the $su$-SSC, these masses are ordered by $0 < \mu_{\hat{\jmath}} < \mu_{\hat{\jmath}+1}$, so we have $\chi > 1$. Hence, $\chi \in (1, \infty)$ parametrizes a set of masses~$\mu_j$ which saturate~\eqref{eq:su-SMC}.

If the masses $\mu_j$ saturate the $su$-SMC, then we can also bound a subset of the couplings. If each $\mu_j$ is given by~\eqref{eq:su-SMC-sol}, then the quantity inside the large parentheses in~\eqref{eq:su-SMC-ineq} is now equal to one. The $k \to \infty$ limit of this expression can be computed using~\eqref{eq:klim-su-4} and yields the following non-linear inequality involving both the masses and couplings:
\begin{align}
    \bigg(
    \frac{ \mu_{j''}^2 }{ \mu_{j\phantom{''}}^2 }
    \bigg)^{(j''-j')(j'-j)}
    \leq
    \frac{
    \big(
    g_{\mfn_{j''},j''}^{2} \, v_{j'',j''}^{(d)}
    \big)^{j'-j}
    \big(
    g_{\mfn_{j},j}^{2} \, v_{j,j}^{(d)}
    \big)^{j''-j'}
    }
    {
    \big(
    g_{\mfn_{j'},j'}^{2} \, v_{j',j'}^{(d)}
    \big)^{j''-j}
    }
    \, .
\end{align}
Choosing $j=\hat{\jmath}$ and $j'=\hat{\jmath}+1$, redefining ${j'' \mapsto j}$, and then using~\eqref{eq:su-SMC-sol} to write the ratio of masses in terms of the parameter $\chi$, this inequality becomes
\begin{align}
\label{eq:su-SMC-gineq}
    g_{\mfn_{j},j}^{2} \,
    v_{j,j}^{(d)}
    \geq
    g_{\mfn_{\hat{\jmath}},\hat{\jmath}}^{2} \,
    v_{\hat{\jmath},\hat{\jmath}}^{(d)} \,
    \chi^{(j-\hat{\jmath})(j-\hat{\jmath}-1)} \,
    \xi^{j-\hat{\jmath}}
    \, ,
\end{align}
where we have defined
\begin{align}
    \xi
    &=
    \frac{ g_{\mfn_{\hat{\jmath}+1},\hat{\jmath}+1}^{2} }
         { g_{\mfn_{\hat{\jmath}},\hat{\jmath}}^{2} \hfill }
    \frac{ v_{\hat{\jmath}+1,\hat{\jmath}+1}^{(d)} }
         { v_{\hat{\jmath},\hat{\jmath}}^{(d)} \hfill }
    \, .
\end{align}
This parameter $\xi > 0$ is strictly positive since the two couplings $g_{\mfn_{\hat{\jmath}},\hat{\jmath}}^{2}$ and $g_{\mfn_{\hat{\jmath}+1},\hat{\jmath}+1}^{2}$ are strictly positive as discussed in~\autoref{sec:intro}. The final inequality~\eqref{eq:su-SMC-gineq} is vacuously true for both ${j = \hat{\jmath}}$ and~${j = \hat{\jmath}+1}$, so it holds for all ${j \geq \hat{\jmath}}$.

We do not know of any theories whose mass spectrum saturates the $su$-SMC, but we can prove by contradiction that such a theory is inconsistent if particles of arbitrarily high spin appear in the weakly-coupled spectrum. If this theory were consistent, then its Wilson coefficients would be finite quantities. Let us consider the Wilson coefficient $a_{1;\hat{\jmath},0}^{(su)}$ using the dispersion relation~\eqref{eq:disprel-su}:
\begin{align}
    a_{1;\hat{\jmath},0}^{(su)}
    &=
    \sum_{n \geq 1}
    \frac{1}{m_n^{2\hat{\jmath}+d-4}}
    \sum_{ j = 0 }^{J_n}
    g_{n,j}^{2}
    +
    \frac{1}{\Lambda^{2\hat{\jmath}+d-4}}
    \int_0^1
    \mathrm{d}z \,
    z^{\hat{\jmath}-1+\frac{d-4}{2}}
    \sum_{j=0}^\infty
    \rho_j \big( \Lambda^2 z^{-1} \big)
    \, .
\end{align}
Here we have used $v_{j,0}^{(d)}=1$. Let us derive a lower bound on this expression. Since $\rho_j(s) \geq 0$ and $g_{n,j}^2 \geq 0$, we can drop the integral and then use~\eqref{eq:subset-su} to restrict the sum over masses from the set of $m_n$ to the subset of $\mu_j$ to find
\begin{align}
    a_{1;\hat{\jmath},0}^{(su)}
    & \geq
    \sum_{j \geq \hat{\jmath}}
    \frac{1}{\mu_j^{2\hat{\jmath}+d-4}} \,
    g_{\mfn_j,j}^{2}
    \, .
\end{align}
Using~\eqref{eq:su-SMC-sol} to write each $\mu_j$ in terms of the parameter $\chi > 1$ and \eqref{eq:su-SMC-gineq} to bound the couplings, we find
\begin{align}
    a_{1;\hat{\jmath},0}^{(su)}
    \geq
    \mu_{\hat{\jmath}}^{-2\hat{\jmath}+4-d} \,
    g_{\mfn_{\hat{\jmath}},\hat{\jmath}}^{2} \, 
    v_{\hat{\jmath},\hat{\jmath}}^{(d)} 
    \sum_{j \geq \hat{\jmath}}
    \chi^{(j-\hat{\jmath})(j-2\hat{\jmath}+1-\frac{d}{2})} \,
    \xi^{j-\hat{\jmath}} \,
    \frac{ 1 }{ v_{j,j}^{(d)} }
    \, .
\end{align}
If the sum over $j$ is infinite (that is, if the theory contains particles of arbitrarily high spin in its weakly-coupled spectrum), then this sum diverges by the ratio test since $\chi > 1$ and $\xi > 0$. In this case, $a_{1;\hat{\jmath},0}^{(su)}$ diverges, and the theory is inconsistent. Therefore, the $su$-SMC cannot be saturated for an infinite set of spins, which implies that the $su$-SMC~\eqref{eq:su-SMC} is not an optimal bound.

\section{The \texorpdfstring{$stu$}{stu}-symmetric case}
\label{sec:stu}

In this section, we prove the Sequential Spin Constraints (SSC) and the Sequential Mass Constraints (SMC) for all $stu$-symmetric amplitudes that satisfy our bootstrap assumptions. Our proof proceeds analogously to the previous section with a few important changes. We repeat much of the language in~\autoref{sec:su} so that this section is self-contained. We again begin by defining a pole-subtracted amplitude and carefully analyzing its Wilson coefficients. Again, our derivation of the dispersion relations for these Wilson coefficients is not new~\cite{Arkani-Hamed:2020blm, Bellazzini:2020cot}, but the constraints on the spectrum (i.e.\ the SSC and SMC) are new to the literature as discussed in \autoref{sec:intro}.

\subsection{The pole-subtracted amplitude}

For any mass $m_n > 0$ in the weakly-coupled spectrum of the $stu$-symmetric amplitude $A^{(stu)}(s,u)$, we define the pole-subtracted amplitude $A^{(stu)}_n(s,u)$ by explicitly subtracting the massless poles and the massive poles with masses $m_p < m_n$:
\begin{align}
\label{eq:polesub-stu}
    A^{(stu)}_n(s,u)
    &=
    A^{(stu)}(s,u)
    {}+{}
    \sum_{\substack{j=0 \\ \mathclap{\text{even}}\mathstrut}}^{J_0}
    \lambda_{j}^2 \, 
    \frac{1}{2}
    \bigg(
      \frac{t^{j}+u^{j}}{s}
    + \frac{u^{j}+s^{j}}{t}
    + \frac{s^{j}+t^{j}}{u}
    \bigg)
\\ \no
    & \quad
    +
    \smash[b]{
    \sum_{p = 1}^{n-1}
    m_p^{6-d}
    \sum_{\substack{j=0 \\ \mathclap{\text{even}}\mathstrut}}^{J_p}
    g_{p,j}^{2} \, 
    \frac{1}{2}
    }
    \bigg\{
    \frac{1}{ \makebox[\widthof{$u$}][c]{$s$}-m_p^2 } \,
    \bigg[
    \mcP_{j}^{(d)} \Big( 1 + \frac{2t}{m_p^2} \Big)
    +
    \mcP_{j}^{(d)} \Big( 1 + \frac{2u}{m_p^2} \Big)
    \bigg]
\\ \no
    & \quad
    \hphantom{
    \sum_{p = 1}^{n-1}
    m_p^{6-d}
    \sum_{\substack{j=0 \\ \mathclap{\text{even}}\mathstrut}}^{J_p}
    g_{p,j}^{2} \, 
    \frac{1}{2}
    \bigg\{
    }
    +
    \frac{1}{ \makebox[\widthof{$u$}][c]{$t$}-m_p^2 } \,
    \bigg[
    \mcP_{j}^{(d)} \Big( 1 + \frac{2u}{m_p^2} \Big)
    +
    \mcP_{j}^{(d)} \Big( 1 + \frac{2s}{m_p^2} \Big)
    \bigg]
\\ \no
    & \quad
    \hphantom{
    \sum_{p = 1}^{n-1}
    m_p^{6-d}
    \sum_{\substack{j=0 \\ \mathclap{\text{even}}\mathstrut}}^{J_p}
    g_{p,j}^{2} \, 
    \frac{1}{2}
    \bigg\{
    }
    +
    \frac{1}{ u-m_p^2 } \,
    \bigg[
    \mcP_{j}^{(d)} \Big( 1 + \frac{2s}{m_p^2} \Big)
    +
    \mcP_{j}^{(d)} \Big( 1 + \frac{2t}{m_p^2} \Big)
    \bigg]
    \bigg\}
    \, ,
\end{align}
with $t=-(s+u)$. The ``subtractions" are accomplished with plus signs because the poles have an overall minus sign as described in~\eqref{eq:resMassive} and~\eqref{eq:resMassless}. We also recall that $J_p$ is the maximum spin exchanged at mass~$m_p$.

This pole-subtracted amplitude is a well-defined functional of the original amplitude since the couplings $\lambda_j^2$ and $g_{p,j}^2$ can be written in terms of $A^{(stu)}(s,u)$ using~\eqref{eq:extractCouplMassless} and~\eqref{eq:extractCouplMassive}. The subtraction manifestly respects the $stu$-symmetry of the original amplitude while eliminating the simple poles in $s$ at ${s=0,m_1^2, \dots, m_{n-1}^2}$ and at~${s=-u,-u{-m_1^2},\dots,-u{-m_{n-1}^2}}$. Otherwise, the pole-subtracted amplitude has the same analytic structure as the original amplitude, depicted in~\autoref{fig:analyticity}.

\subsubsection{Polynomial bound}

Due to the various powers of~$s$ in the numerators, the pole-subtracted amplitude~\eqref{eq:polesub-stu} is polynomially-bounded by
\begin{align}
    \lim_{\substack{ |s| \to \infty \\
                     \mathclap{\text{fixed } u<0 }}}
    \frac{A^{(stu)}_n(s,u)}
         {s^{\mcJ_n}}
    = 0
\end{align}
with
\begin{align}
\label{eq:Jn-stu}
    \mcJ_n 
    =
    \max(\alpha-1, J_0, J_1, \dots, J_{n-1})+1
    \, ,
\end{align}
where $\alpha$ is the degree of the polynomial bound of the original amplitude $A^{(stu)}(s,u)$ in~\eqref{eq:bound}. This polynomial bound $\mcJ_n$ is finite since each maximum spin $J_n$ is finite (by assumption). The $J_n$ must be even as a consequence of $stu$-symmetry, as discussed below~\eqref{eq:partial}. If there are no massless poles, we set $J_0 = -2$ by convention.

\subsubsection{Low-energy expansion}

Since we have subtracted the massless poles, the low-energy expansion near $s=u=0$ is analytic in $s$ and $u$ and given by
\begin{align}
\label{eq:lowenergy-stu}
    A^{(stu)}_n(s,u) 
    &=
    \sum_{k = 0}^\infty
    \sum_{q = 0}^k
    a^{(stu)}_{n;k,q} \,
    s^{k-q} u^q
    \, .
\end{align}
This low-energy expansion is valid for $|s|, |t|, |u| < m_n^2$. As a consequence of $stu$-symmetry, the Wilson coefficients obey homogeneous linear equations, graded by the Mandelstam order~$k$. For instance, the subset of $su$-symmetry implies $a^{(stu)}_{n;k,q} = a^{(stu)}_{n;k,k-q}$ for all $k \geq q \geq 0$.

\subsubsection{Dispersion relations}

We can derive a dispersion relation for the the Wilson coefficient $a^{(stu)}_{n;k,q}$ with $k-q \geq \mcJ_n$ by considering the following contour integral at fixed $u \in (-\frac{1}{2}m_n^2, 0)$:
\begin{align}
\label{eq:contour-stu}
    \Bigg\{
    \quad
    \ointctrclockwise\limits_{\mathclap{s=0^{\phantom{2}}}}
    \quad \,
    +
    \sum_{m_p \geq m_n}
    \Big(
    \,
    \ointctrclockwise\limits_{\mathclap{s=m_p^2}}
    \quad \,
    +
    \quad \,
    \ointctrclockwise\limits_{\mathclap{s=-u-m_p^2}}
    \,
    \Big)
    \quad \,
    +
    \quad \,
    \ointctrclockwise\limits_{\mathclap{s \in [\Lambda^2,\infty] }}
    \quad \,
    +
    \quad \,
    \ointctrclockwise\limits_{\mathclap{s \in [-u-\Lambda^2,\infty] }}
    \quad \,
    +
    \quad \,
    \varointclockwise\limits_{\mathclap{s=\infty^{\phantom{2}}}}
    \quad
    \Bigg\}
    &
\no \\
    {} \times {}
    \frac{ \mathrm{d}s }{ 2\pi i }
    \frac{ A^{(stu)}_n(s,u) }{ s^{k-q+1} }
    &=
    0
    \, .
\end{align}
The full contour is depicted in~\autoref{fig:contour-stu} and trivially vanishes by Cauchy's integral theorem.
\begin{itemize}
\item The contour around $s = 0$ extracts the Wilson coefficients in~\eqref{eq:lowenergy-stu}, which can be used since $|s| < \half m_n^2$ and $u \in (-\frac{1}{2}m_n^2, 0)$ imply that $|s|, |u|, |s+u| < m_n^2$:
\begin{align}
\label{eq:contour-stu1}
    \ointctrclockwise\limits_{\mathclap{s=0}}
    \,
    \frac{ \mathrm{d}s }{ 2\pi i }
    \frac{ A^{(stu)}_n(s,u) }{ s^{k-q+1} }
    &=
    \sum_{q' = 0}^\infty
    a^{(stu)}_{n;k-q+q',q'} \,
    u^{q'}
    \, .
\intertext{\item The contours around each $s=m_p^2$ can be evaluated using~\eqref{eq:resMassive}:}
\label{eq:contour-stu2}
    \ointctrclockwise\limits_{\mathclap{s=m_p^2}}
    \,
    \frac{ \mathrm{d}s }{ 2\pi i }
    \frac{ A^{(stu)}_n(s,u) }{ s^{k-q+1} }
    &=
    -
    \frac{1}{m_p^{2k-2q+d-4}}
    \sum_{\substack{j=0 \\ \mathclap{\text{even}}\mathstrut}}^{J_p}
    g_{p,j}^{2} \,
    \mcP_{j}^{(d)} \Big( 1 + \frac{2u}{m_p^2} \Big)
    \, .
\intertext{\item The contours around each $s=-u-m_p^2$ can be similarly evaluated by first using the $st$-symmetry relation ${A^{(stu)}_n(s,u) = A^{(stu)}_n(-s{-u},u)}$ and then using~\eqref{eq:resMassive}:}
\label{eq:contour-stu3}
    \ointctrclockwise\limits_{\mathclap{s=-u-m_p^2}}
    \,
    \frac{ \mathrm{d}s }{ 2\pi i }
    \frac{ A^{(stu)}_n(s,u) }{ s^{k-q+1} }
    &=
    -
    \frac{ 1 }{ m_p^{2k-2q+d-4} }
    \frac{ (-)^{k-q} }
         { \big( 1 + \frac{u \mathstrut}{m_p^2} \big)^{k-q+1} }
    \sum_{\substack{j=0 \\ \mathclap{\text{even}}\mathstrut}}^{J_p}
    g_{p,j}^{2} \,
    \mcP_{j}^{(d)} \Big( 1 + \frac{2u \mathstrut}{m_p^2} \Big)
    \, .
\intertext{\item The contour around $s \in [\Lambda^2, \infty]$ computes the discontinuity of the integrand across the real $s$-axis, and within this integral, the imaginary part of the pole-subtracted amplitude is equal to the imaginary part of the original amplitude $A^{(stu)}(s,u)$, which can be expanded as a sum over the spectral density functions $\rho_j(s)$ defined in~\eqref{eq:rho}:}
\label{eq:contour-stu4}
    \ointctrclockwise\limits_{\mathclap{s \in [\Lambda^2,\infty] }}
    \,
    \frac{ \mathrm{d}s }{ 2\pi i }
    \frac{ A^{(stu)}_n(s,u) }{ s^{k-q+1} }
    &=
    -
    \int_{\Lambda^2}^\infty
    \frac{ \mathrm{d}s }{ s^{k-q+1} }
    \sum_{\substack{j = 0 \\ \mathclap{\text{even}}\mathstrut}}^\infty
    \rho_j(s) \,
    \mcP_j^{(d)} \Big( 1 + \frac{2u}{s} \Big)
    \vphantom{\frac{ \Im A^{(su)}(s,u) }{ s^{k-q+1} }}
    \, .
\intertext{\item The contour around $s \in [\infty,-u-\Lambda^2]$ can be similarly evaluated using $st$-symmetry:}
\label{eq:contour-stu5}
    \ointctrclockwise\limits_{\mathclap{s \in [-u-\Lambda^2,\infty] }}
    \,
    \frac{ \mathrm{d}s }{ 2\pi i }
    \frac{ A^{(stu)}_n(s,u) }{ s^{k-q+1} }
    &=
    -
    \int_{\Lambda^2}^\infty
    \frac{ \mathrm{d}s }{ s^{k-q+1} }
    \frac{ (-)^{k-q} }
         { \big( 1 + \frac{u \mathstrut}{ s \mathstrut } \big)^{k-q+1} }
    \sum_{\substack{j = 0 \\ \mathclap{\text{even}}\mathstrut}}^\infty
    \rho_j(s) \,
    \mcP_{j}^{(d)} \Big( 1 + \frac{2u \mathstrut}{s \mathstrut} \Big)
    \, .
\end{align}
\item Finally, the contour at infinity vanishes since $k-q \geq \mcJ_n$.
\end{itemize}
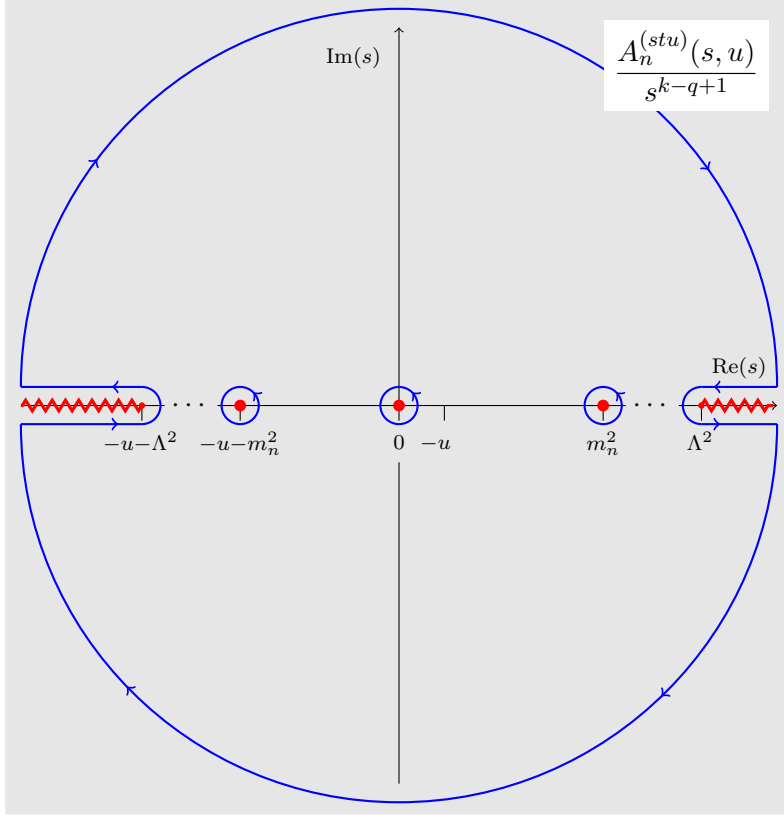
\begin{figure}
\centering
\begin{tikzpicture}
    \clip (-5.5,-5.5) rectangle (5.5,5.5);
    
    \draw[fill=gray!20, color=gray!20]
         (-5.2,-5.4) --
         (-5.2, 5.4) --
         ( 5.2, 5.4) --
         ( 5.2,-5.4) -- cycle;
    
    \coordinate (m0) at ( 0.0, 0.0);
    \coordinate (m1) at ( 1.4, 0.0);
    \coordinate (mn) at ( 2.7, 0.0);
    \coordinate (mL) at ( 4.0, 0.0);

    \coordinate (u)  at (-0.6, 0.0);

    \coordinate (u0) at ($-1*(u)-(m0)$);
    \coordinate (u1) at ($-1*(u)-(m1)$);
    \coordinate (un) at ($-1*(u)-(mn)$);
    \coordinate (uL) at ($-1*(u)-(mL)$);

    \node at ( 4.5, 0.5) {$\scriptstyle\Re(s)$};
    \node at (-0.6, 4.6) {$\scriptstyle\Im(s)$};

    \coordinate (xr) at ( 0.3, 0.0);
    \coordinate (yr) at ( 0.0, 0.2);
    
    \draw     (-5.0, 0.0)   -- ($(uL)+(xr)$);
    \draw     ($(un)-(xr)$) -- ($(mn)+(xr)$);
    \draw[->] ($(mL)-(xr)$) -- ( 5.0, 0.0);

    \draw     ( 0.0,-5.0)   -- ( 0.0,-0.75);
    \draw[->] ( 0.0, 0.0)   -- ( 0.0, 5.0);

    \draw (m0) -- ($(m0)-(yr)$);
    \draw (mn) -- ($(mn)-(yr)$);
    \draw (mL) -- ($(mL)-(yr)$);

    \draw (u0) -- ($(u0)-(yr)$);
    \draw (un) -- ($(un)-(yr)$);
    \draw (uL) -- ($(uL)-(yr)$);
    
    \node at ($(m0)-2.5*(yr)$) {$\scriptstyle 0 \mathstrut$};
    \node at ($(mn)-2.5*(yr)$) {$\scriptstyle m_n^2$};
    \node at ($(mL)-2.5*(yr)$) {$\scriptstyle \Lambda_{\phantom{n}}^2$};

    \node at ($(u0)-2.5*(yr)$) {$\scriptstyle -u \phantom{-}$};
    \node at ($(un)-2.5*(yr)$) {$\scriptstyle -u-m_n^2$};
    \node at ($(uL)-2.5*(yr)$) {$\scriptstyle -u-\Lambda_{\phantom{n}}^2$};

    \node at ($0.5*(mL)+0.5*(mn)$) {$\cdots$};
    \node at ($0.5*(uL)+0.5*(un)$) {$\cdots$};

    \filldraw[color=red] (m0) circle (2pt);
    \filldraw[color=red] (mn) circle (2pt);
    \filldraw[color=red] (mL) circle (1pt);

    \filldraw[color=red] (un) circle (2pt);
    \filldraw[color=red] (uL) circle (1pt);

    \draw[color=red, very thick, decorate,
          decoration={zigzag, amplitude=2pt, segment length=5pt}]
        (mL) -- ( 4.95, 0.0);   
        
    \draw[color=red, very thick, decorate,
          decoration={zigzag, amplitude=2pt, segment length=5pt}]
        (uL) -- (-5.0, 0.0);   

    \draw[color=blue, thick, postaction={decorate},
          decoration={markings, mark=at position 0.125 with {\arrow{>}}}]
         (m0) circle [radius=7pt];
    \draw[color=blue, thick, postaction={decorate},
          decoration={markings, mark=at position 0.125 with {\arrow{>}}}]
         (mn) circle [radius=7pt];
    \draw[color=blue, thick, postaction={decorate},
          decoration={markings, mark=at position 0.125 with {\arrow{>}}}]
         (un) circle [radius=7pt];

    \draw[color=blue, thick] ($(mL)+(0,7pt)$) arc ( 90:270:7pt);
    \draw[color=blue, thick] ($(uL)-(0,7pt)$) arc (-90: 90:7pt);

    \draw[color=blue, thick, postaction={decorate},
          decoration={markings, mark=at position 0.25 with {\arrow{<}}}]
         ($(mL)+(0,7pt)$) -- ($(5,0)+(0,7pt)$);
    \draw[color=blue, thick, postaction={decorate},
          decoration={markings, mark=at position 0.25 with {\arrow{>}}}]
         ($(mL)-(0,7pt)$) -- ($(5,0)-(0,7pt)$);
    \draw[color=blue, thick, postaction={decorate},
          decoration={markings, mark=at position 0.25 with {\arrow{>}}}]
         ($(uL)+(0,7pt)$) -- ($(-5,0)+(0,7pt)$);
    \draw[color=blue, thick, postaction={decorate},
          decoration={markings, mark=at position 0.25 with {\arrow{<}}}]
         ($(uL)-(0,7pt)$) -- ($(-5,0)-(0,7pt)$);

    \draw[color=blue, thick, postaction={decorate},
          decoration={markings, mark=at position 0.2 with {\arrow{<}},
                                mark=at position 0.8 with {\arrow{<}}}]
         ($(5,0)+(0,7pt)$) arc (0:180:5);
    \draw[color=blue, thick, postaction={decorate},
          decoration={markings, mark=at position 0.25 with {\arrow{<}},
                                mark=at position 0.75 with {\arrow{<}}}]
         ($(-5,0)-(0,7pt)$) arc (-180:0:5);

    \node[fill=white] at ( 3.8, 4.5)
        {$\displaystyle\frac{A^{(stu)}_n(s,u)}{s^{k-q+1}}$};
    
\end{tikzpicture}

\caption{The contour integral in~\eqref{eq:contour-stu}.}

\label{fig:contour-stu}

\end{figure}
Each contribution to the full contour integral is a power series in~$u$, valid for all values of ${u \in (-\half m_n^2, 0)}$ with no obstruction to taking the limit ${u \to 0}$. Thus, we can act with the operator $\frac{1}{q!} (\partial / \partial u)^q$ on the full contour integral before taking ${u \to 0}$ to extract the coefficient of $u^q$.

After some straightforward algebra and a change of integration variables to the dimensionless combination ${z = \Lambda^2 s^{-1}}$, we arrive at the following dispersion relation for the Wilson coefficients of $A_n^{(stu)}(s,u)$. This relation holds for all integers $n \geq 1$ with $m_n$ in the weakly-coupled spectrum and for all integers $k,q \geq 0$ satisfying $k-q \geq \mcJ_n$:
\begin{align}
\label{eq:disprel-stu}
    a^{(stu)}_{n;k,q}
    &=
    \sum_{m_p \geq m_n}
    \frac{1}{m_p^{2k+d-4}}
    \sum_{\substack{j=0 \\ \mathclap{\text{even}}\mathstrut}}^{J_p}
    g_{p,j}^{2} \,
    w_{j;k,q}^{(d)}
\no \\
    & \quad
    +
    \frac{1}{\Lambda^{2k+d-4}}
    \int_0^1
    \mathrm{d}z \,
    z^{k-1+\frac{d-4}{2}}
    \sum_{\substack{j=0 \\ \mathclap{\text{even}}\mathstrut}}^\infty
    \rho_j \big( \Lambda^2 z^{-1} \big) \,
    w_{j;k,q}^{(d)}
    \, .
\end{align}
Here $w_{j;k,q}^{(d)}$ is a sign-indefinite linear combination of the $v_{j,q}^{(d)}$ coefficients, defined by 
\begin{align}
\label{eq:wjkqd}
    w_{j;k,q}^{(d)}
    =
    v_{j,q}^{(d)}
    +
    \sum_{q'=0}^{q}
    \binom{k-q'}{q-q'}
    (-)^{k-q'}
    v_{j,q'}^{(d)}
    \, .
\end{align}
An explicit expression for $v_{j,q}^{(d)}$ is given in~\eqref{eq:Gegv}. The sum over $q'$ in~\eqref{eq:wjkqd} can actually be performed, resulting in a ${}_3F_2$ hypergeometric function:
\begin{align}
\label{eq:wjkqd-2}
    w_{j;k,q}^{(d)}
    =
    v_{j,q}^{(d)}
    +
    (-)^k
    \binom{k}{q} \,
    {{}_3F_2}
    \Big[
    \begin{smallmatrix}
    -j, \,\, j+d-3, \,\, -q
    \\
    \frac{1}{2}(d-2), \,\, -k
    \end{smallmatrix}
    \,;\, 1
    \Big]
    \, .
\end{align}
This hypergeometric function is a well-defined finite series, but one must take care when expressing it using the series representation of ${}_3F_2$ because of the negative ``downstairs" argument $-k$. In any case, we can define the following combination of these ${}_3F_2$'s which is anti-symmetric under $q \leftrightarrow k-q$:
\begin{align}
\label{eq:fjkqd}
    f_{j;k,q}^{(d)}
    &=
    {{}_3F_2}
    \Big[
    \begin{smallmatrix}
    -j, \,\, j+d-3, \,\, -q
    \\
    \frac{1}{2}(d-2), \,\, -k
    \end{smallmatrix}
    \,;\, 1
    \Big]
    -
    {{}_3F_2}
    \Big[
    \begin{smallmatrix}
    -j, \,\, j+d-3, \,\, -k+q
    \\
    \frac{1}{2}(d-2), \,\, -k
    \end{smallmatrix}
    \,;\, 1
    \Big]
    \, .
\end{align}
This $f$ coefficient vanishes for all non-negative integers $j,k,q$ with $j$ even and $k \geq j, q$. While we do not have an analytic proof of this property, we have numerically verified it for all values of $j,k,q$ up to $k = 50$ and for a random selection of $j,k,q$ with ${51 \leq k \leq 200}$.

In contrast to the $su$-symmetric dispersion relation~\eqref{eq:disprel-su} for $a^{(su)}_{n;k,q}$ where the sums over $j$ begin with $j=q$, the sums in the $stu$-symmetric dispersion relation~\eqref{eq:disprel-stu} begin at $j=0$ because the $w$ coefficients are generically non-zero for all $j \geq 0$. Hence, all even spins contribute to the $stu$-symmetric Wilson coefficients $a^{(stu)}_{n;k,q}$. Moreover, since the $w$ coefficients are sign-indefinite, so to are the $stu$-symmetric Wilson coefficients, again in contrast to the $su$-symmetric Wilson coefficients which are manifestly non-negative.

\subsubsection{The \texorpdfstring{$b$}{b} coefficients}

To proceed, it will be useful to have a quantity analogous to the $su$-symmetric Wilson coefficients which is manifestly non-negative and which depends only on a restricted set of spins. To that end, we define the $b$ coefficients by replacing each $w$ coefficient in the dispersion relation for $a^{(stu)}_{n;k,q}$ with the corresponding $v$ coefficient:
\begin{align}
\label{eq:disprel-stu-b}
    b_{n;k,q}
    &=
    \sum_{m_p \geq m_n}
    \frac{1}{m_p^{2k+d-4}}
    \sum_{\substack{j=2\ceil{q/2} \\ \mathclap{\text{even}}\mathstrut}}^{J_p}
    g_{p,j}^{2} \,
    v_{j,q}^{(d)}
\no \\
    & \quad
    +
    \frac{1}{\Lambda^{2k+d-4}}
    \int_0^1
    \mathrm{d}z \,
    z^{k-1+\frac{d-4}{2}}
    \sum_{\substack{j=2\ceil{q/2} \\ \mathclap{\text{even}}\mathstrut}}^\infty
    \rho_j \big( \Lambda^2 z^{-1} \big) \,
    v_{j,q}^{(d)}
\end{align}
Each $b$ coefficient is manifestly positive, and the sums over $j$ begin with $j=2\ceil{q/2}$, the smallest even integer greater than or equal to $q$. The $a$ and $b$ coefficients can be related using the definition~\eqref{eq:wjkqd} of the $w$ coefficients in terms of the $v$ coefficients. Clearly, each Wilson coefficient is a linear combination of the $b$ coefficients.

Now, the $su$-symmetry relation $A_n^{(stu)}(s,u) = A_n^{(stu)}(u,s)$ implies that the Wilson coefficients obey $a^{(stu)}_{n;k,q} = a^{(stu)}_{n;k,k-q}$. If both ${k-q \geq \mcJ_n}$ and ${q \geq \mcJ_n}$, then this crossing symmetry relation can be written in terms of the $b$ coefficients using equations~\eqref{eq:disprel-stu}, \eqref{eq:wjkqd-2}, \eqref{eq:fjkqd}, and~\eqref{eq:disprel-stu-b}:
\begin{align}
\label{eq:b-crossing}
    \Delta b_{n;k,q}
    &=
    b_{n;k,k-q} - b_{n;k,q}
    \smash[t]{\vphantom{ \Bigg\{ }}
\no \\
    &=
    \underbrace{
    a^{(stu)}_{n;k,k-q}
    - a^{(stu)}_{n;k,q}
    \vphantom{\bigg|}
        }_{\color{red} 0 \mathstrut}
    {}+{}
    \sum_{q'=0}^{q}
    \binom{k-q'}{q-q'}
    (-)^{k-q'}
    b_{n;k,q'}
    -
    \sum_{q'=0}^{k-q}
    \binom{k-q'}{k-q-q'}
    (-)^{k-q'}
    b_{n;k,q'}
\no \\
    &=
    (-)^k
    \binom{k}{q}
    \Bigg\{
    \,
    \sum_{m_p \geq m_n}
    \frac{1}{m_p^{2k+d-4}}
    \sum_{\substack{j=2\ceil{k/2} \\ \mathclap{\text{even}}\mathstrut}}^{J_p}
    g_{p,j \vphantom{k}}^{2} \,
    f_{j;k,q}^{(d)}
\no \\
    & \quad 
    \hphantom{ (-)^k \binom{k}{q} \Bigg\{ }
    {}+{}
    \frac{1}{\Lambda^{2k+d-4}}
    \int_0^1
    \mathrm{d}z \,
    z^{k-1+\frac{d-4}{2}}
    \sum_{\substack{j=2\ceil{k/2} \\ \mathclap{\text{even}}\mathstrut}}^\infty
    \rho_j \big( \Lambda^2 z^{-1} \big) \,
    f_{j;k,q}^{(d)}
    \, 
    \Bigg\}
    \, ,
\end{align}
where we have defined $\Delta b_{n;k,q} = b_{n;k,k-q} - b_{n;k,q}$. The sums over $j$ in this expression begin with $j=2\ceil{k/2}$ because the $f$ coefficients vanish for even $j \leq k$ as discussed above.

\subsubsection{The \texorpdfstring{$k \to \infty$}{k to infinity} limit}

Following the strategy of the previous section, let us now examine the $k \to \infty$ limit of these dispersion relations with $q$ held fixed. Because of the complicated $k$-dependence of the~$w$ coefficients, we cannot simply compute the infinite $k$ limit of the $stu$-symmetric Wilson coefficients. We can, however, compute this limit for the $b$ coefficients using~\eqref{eq:disprel-stu-b}. As before, we must first eliminate the overall mass dimension in this equation by multiplying by an appropriate power of some mass $m$. The particular choice of this mass then determines whether the limit is zero, non-zero and finite, or infinite.
\begin{itemize}
\item If $m < m_{n}$, then the limit vanishes:
\begin{align}
\label{eq:klim-stu-1}
    \lim_{k \to \infty}
    \Big(
    m^{2k+d-4} \, b_{n;k,q}
    \Big)
    &=
    0
    \, .
\end{align}
\item If $m = m_{n}$, then the limit results in a finite sum over the spin-$j$ couplings at that mass level beginning with $j=2\ceil{q/2}$,
\begin{align}
\label{eq:klim-stu-2}
    \lim_{k \to \infty}
    \Big(
    m_{n}^{2k+d-4} \, b_{n;k,q}
    \Big)
    &=
    \sum_{\substack{j=2\ceil{q/2} \\ \mathclap{\text{even}}\mathstrut}}^{J_n}
    g_{n,j}^{2} \, 
    v_{j,q}^{(d)}
    \, ,
\end{align}
which vanishes if $q > J_n$. This sum is finite because (by assumption) the amplitude exchanges a finite set of spins with maximum spin $J_n$ at mass $m_n$.
\item If $m > m_{n}$, then the limit generically diverges, but if $m = \mu_{2\ceil{q/2}}$ is the first mass level at which a spin-$(2\ceil{q/2})$ or higher state is exchanged (with $\mu_q > m_n$, assuming that such a mass level exists), then the limit results in a finite sum over the spin\nobreakdash-$j$ couplings at that mass level beginning with $j=2\ceil{q/2}$, similar to the previous equation:
\begin{align}
\label{eq:klim-stu-3}
    \lim_{k \to \infty}
    \Big(
    \mu_{2\ceil{q/2}}^{2k+d-4} \, b_{n;k,q}
    \Big)
    &=
    \sum_{\substack{j=2\ceil{q/2} \\ \mathclap{\text{even}}\mathstrut}}
        ^{ J_{\mfn_{2\ceil{q/2}}} }
    g_{\mfn_{2\ceil{q/2}},j}^{2} \, 
    v_{j,q}^{(d)}
    \, ,
\end{align}
where $\mfn_{2\ceil{q/2}}$ is the index of the first mass level at which a spin-$(2\ceil{q/2})$ state is exchanged so that ${m_{\mfn_{2\ceil{q/2}}} = \mu_{2\ceil{q/2}}}$. This limit is finite because, in this case, no terms proportional to $x^k$ with ${x > 1}$ appear after multiplying the dispersion relation~\eqref{eq:disprel-stu-b} by $\mu_{2\ceil{q/2}}^{2k+d-4}$.
\end{itemize}

Let us also analyze the $k \to \infty$ limit of the crossing symmetry relation~\eqref{eq:b-crossing} for the $b$ coefficients. To compute these limits, we need a few intermediate results. First, we can use the series representation of the hypergeometric function to verify that the $k \to \infty$ limit of the $f$ coefficient~\eqref{eq:fjkqd} vanishes with fixed $j,q,d$:
\begin{align}
\label{eq:klim-stu-int1}
    \lim_{k \to \infty}
    f_{j;k,q}^{(d)}
    =
    0
    \, .
\end{align} 
Also, for $|x|<1$ and fixed $q$,
\begin{align}
\label{eq:klim-stu-int2}
    \lim_{k \to \infty}
    x^k
    \binom{k}{q}
    =
    0
\end{align}
since $\binom{k}{q} \sim k^q$ grows polynomially with $k$. With these results in hand, we can compute the $k \to \infty$ limit of~\eqref{eq:b-crossing} multiplied by $m^{2k+d-4}$.
\begin{itemize}
\item If $m < m_{n}$, then the limit trivially vanishes as a consequence of~\eqref{eq:klim-stu-int1} and~\eqref{eq:klim-stu-int2}:
\begin{align}
\label{eq:klim-stu-4}
    \lim_{k \to \infty}
    \Big(
    m^{2k+d-4} \, \Delta b_{n;k,q}
    \Big)
    &=
    0
    \, .
\end{align}
\item If $m = m_r $ with $m_r \geq m_n$ is any mass in the weakly-coupled spectrum of $A^{(su)}_n(s,u)$, then the limit requires some care. The terms proportional to $(m_r / m_p)^{2k}$ with ${m_p > m_r}$ or to $(m_r / \Lambda)^{2k}$ trivially vanish as a consequence of~\eqref{eq:klim-stu-int1} and~\eqref{eq:klim-stu-int2}. This leaves a finite number of terms proportional to $(m_r / m_p)^{2k}$ with ${m_p \leq m_r}$. The $k \to \infty$ limit of these terms appears to diverge, but they in fact vanish because the set of~$j$ satisfying the summation criterion $2\ceil{k/2} \leq j \leq J_p$ is empty in the limit $k \to \infty$ since each $J_p$ is finite by assumption. Thus,
\begin{align}
\label{eq:klim-stu-5}
    \lim_{k \to \infty}
    \Big(
    m_r^{2k+d-4} \, \Delta b_{n;k,q}
    \Big)
    &=
    0
    \, .
\end{align}
\end{itemize}
We conclude that the infinite $k$ limit of $m^{2k+d-4} \, \Delta b_{n;k,q}$ vanishes for any $m$ in the weakly-coupled spectrum.

\subsubsection{An inequality from positivity}

Again in close analogy to the previous section, we can use positivity to write an inequality relating the $b$ coefficients with different values of the ``$k$" subscript. For any integer $q \geq 0$ and any mass $m$, the following ratios yield a sequence of the form discussed in~\autoref{sec:su}:
\begin{align}
\label{eq:Ai-stu}
    A_i
    =
    m^{2i} \,
    \frac{ b_{n;q+\mcJ_n+i,q} }
         { b_{n;q+\mcJ_n,q \phantom{+i}} }
    \, .
\end{align}
For any integers $i_2 \geq i_1 \geq 0$, positivity (and Jensen's theorem) implies $A_{i_1}^{i_2} \leq A_{i_2}^{i_1}$. If we choose $q=k-\mcJ_n-i_1$ for any integer $k \geq 2\mcJ_n+i_1$, then this inequality becomes
\begin{align}
    \big(
    b_{n;k,k-\mcJ_n-i_1}
    \big)^{i_2}
    &
    \leq
    \big(
    b_{n;k+i_2-i_1,k-\mcJ_n-i_1}
    \big)^{i_1}
    \big(
    b_{n;k-i_1,k-\mcJ_n-i_1}
    \big)^{i_2-i_1}
    \, ,
\end{align}
or equivalently
\begin{align}
\label{eq:ineq-stu}
    \big(
    b_{n;k,\mcJ_n+i_1}
    + \Delta b_{n;k,\mcJ_n+i_1}
    \big)^{i_2}
    \leq
    \big(
    b_{n;k+i_2-i_1,\mcJ_n+i_2}
    &
    {} + \Delta b_{n;k+i_2-i_1,\mcJ_n+i_2}
    \big)^{i_1}
\no \\
    \times
    \big(
    b_{n;k-i_1,\mcJ_n}
    &
    {} + \Delta b_{n;k-i_1,\mcJ_n}
    \big)^{i_2-i_1}
    \, ,
\end{align}
where $\Delta b_{n;k,q} = b_{n;k,k-q}-b_{n;k,q}$. Each term which appears in this final inequality has a dispersion relation and a convergent $k \to \infty$ limit.

\subsection{Proving the \texorpdfstring{$stu$}{stu}-SSC and \texorpdfstring{$stu$}{stu}-SMC}

Both the $stu$-symmetric SSC and SMC follow directly from the inequality~\eqref{eq:ineq-stu}. For each proof, we simply multiply both sides of the inequality by an appropriate factor and then analyze the $k \to \infty$ limit.

\subsubsection{Proving the \texorpdfstring{$stu$}{stu}-SSC}

Let us first demonstrate that the $stu$-SSC can be written as $J_n \leq 2\ceil{\mcJ_n/2}$, where $2\ceil{\mcJ_n/2}$ is the smallest even integer greater than or equal to $\mcJ_n$, the polynomial bound in~\eqref{eq:Jn-su}. The ceiling function is monotonic, so it commutes with the maximum function. Moreover, each maximum spin $J_p$ is even, so ${2\ceil{(J_p+1)/2} = J_p+2}$. Finally, for all integers $\alpha$, we have ${2\ceil{\alpha/2} = 2\floor{(\alpha-1)/2}+2}$. Thus,
\begin{align}
    2\ceil{\mcJ_n/2}
    &=
    \max(2\floor{(\alpha-1)/2}, J_0, J_1, \dots, J_{n-1})+2
    \, ,
\end{align}
so the $stu$-SSC can be written as $J_n \leq 2\ceil{\mcJ_n/2}$ as claimed.

We shall prove this statement by contradiction (mirroring our proof of the $su$-SSC). Let us first assume that ${J_n \geq 2\ceil{\mcJ_n/2} + 2}$. Next we consider the inequality~\eqref{eq:ineq-stu} and multiply both sides by $m_n^{(2k+d-4)i_2}$. We then set ${i_1 = J_n - 2\ceil{\mcJ_n/2}}$ and ${i_2 = i_1 + 2}$ so that both $i_1$ and $i_2$ are even and ${i_2 > i_1 \geq 2}$. Now we can compute the $k \to \infty$ limit of this expression using~\eqref{eq:klim-stu-2} and~\eqref{eq:klim-stu-5}. The $\Delta b$ terms vanish in this limit while the $b$ terms yield finite sums over $j$ as follows:
\begin{align}
    \Bigg(
    \sum_{\substack{ j = J_n \\ \mathclap{\text{even}}\mathstrut}}^{ J_n }
    g_{n,j}^{2} \, 
    v_{j,J_n +\mcJ_n- 2\ceil{\mcJ_n/2}}^{(d)}
    \Bigg)^{ J_n - 2\ceil{\mcJ_n/2} + 2 }
    &\leq
    \Bigg(
    \sum_{\substack{ j = J_n+2 \\ \mathclap{\text{even}}\mathstrut}}^{ J_n }
    g_{n,j}^{2} \, 
    v_{j,J_n +\mcJ_n- 2\ceil{\mcJ_n/2}+2}^{(d)}
    \Bigg)^{ J_n - 2\ceil{\mcJ_n/2} }
\no \\
    & \quad \times
    \Bigg(
    \sum_{\substack{ j = 2\ceil{\mcJ_n/2} \\ \mathclap{\text{even}}\mathstrut}}^{ J_n }
    g_{n,j}^{2} \, 
    v_{j,\mcJ_n}^{(d)}
    \Bigg)^2
    \, .
\end{align}
The first sum on the right-hand side vanishes identically because the lower limit is larger than the upper limit, and the sum on the left-hand side is manifestly non-negative. The single $v$ coefficient which appears on the left-hand side is either $v_{J_n,J_n}^{(d)}$ or $v_{J_n,J_n-1}^{(d)}$, depending on whether $\mcJ_n$ is even or odd. Both of these $v$ coefficients are strictly positive, so we must have $g_{n,J_n}^2 = 0$, which contradicts our bootstrap assumptions. Recall that in~\autoref{sec:intro} we assumed that the coupling of the maximum spin at each mass level $g_{n,J_n}^2 > 0$ was strictly positive without loss of generality. Having reached a contradiction, we conclude that $J_n \leq 2\ceil{\mcJ_n/2} + 1$, but since $J_n$ must be even, we have $J_n \leq 2\ceil{\mcJ_n/2}$. This completes our proof of the $stu$-symmetric SSC.

\subsubsection{Proving the \texorpdfstring{$stu$}{stu}-SMC}

To prove the $stu$-SMC, we begin with a corollary of the $stu$-SSC. If $\mfn_j$ (with $j$ even) is the index of the first mass level at which a spin-$j$ state is exchanged (so that $m_{\mfn_j} = \mu_j$), then the $stu$-SSC implies that the maximum spin at this mass level is equal to $j$. More precisely, $J_{\mfn_j} = j$ and $\mcJ_{\mfn_j} = j$ for all even ${j \geq \max(2\ceil{\alpha/2},J_0+2)}$ with $\mu_j$ in the weakly-coupled spectrum. In other words, the first appearance of a spin-$j$ state must be at a mass level at which it is the maximum spin.

This corollary implies that the limit~\eqref{eq:klim-stu-3} with ${q = j}$ for even ${j \geq \max(2\ceil{\alpha/2},J_0+2)}$ such that $\mu_j \geq m_n$ simplifies to
\begin{align}
\label{eq:klim-stu-6}
    \lim_{k \to \infty}
    \Big(
    \mu_j^{2k+d-4} \, b_{n;k,j}
    \Big)
    &=
    g_{\mfn_j,j}^{2} \, 
    v_{j,j}^{(d)}
    > 0
    \, ,
\end{align}
which is strictly positive because both $v_{j,j}^{(d)} > 0$ and $g_{\mfn_j,j}^{2} = g_{\mfn_j,J_{\mfn_j}}^2 > 0$ are positive.

Now we consider the inequality~\eqref{eq:ineq-stu}, setting ${n = \mfn_j}$ for even ${j \geq \max(2\ceil{\alpha/2},J_0+2)}$ with $\mu_j$ in the weakly-coupled spectrum so that $\mcJ_n = j$. We restrict to even ${i_2 \geq i_1 \geq 0}$ and define ${j' = j+i_1}$ and ${j'' = j+i_2}$ with $\mu_{j'}, \mu_{j''}$ also in the weakly-coupled spectrum. After some rearranging,~\eqref{eq:ineq-stu} becomes
\begin{align}
    1 \leq
    \frac{
    \big(
    b_{\mfn_j;k+j''-j',j''}
    + \Delta b_{\mfn_j;k+j''-j',j''}
    \big)^{j'-j}
    \big(
    b_{\mfn_j;k+j-j',j}
    + \Delta b_{\mfn_j;k+j-j',j}
    \big)^{j''-j'}
    }
    {
    \big(
    b_{\mfn_j;k,j'}
    + \Delta b_{\mfn_j;k,j'}
    \big)^{j''-j}
    }
    \, .
\end{align}
Next we multiply both sides of this expression by
\begin{align}
    \frac{
    \big(
    \mu_{j''}^{2(k+j''-j')+d-4}
    \big)^{j'-j}
    \big(
    \mu_{j \vphantom{j'}}^{2(k+j-j')+d-4}
    \big)^{j''-j'}
    }
    {
    \big(
    \mu_{j'}^{2k+d-4}
    \big)^{j''-j}
    }
\end{align}
to find a horribly messy inequality,
\begin{align}
\label{eq:stu-SMC-ineq}
    & \quad
    \bigg(
    \frac{ \mu_{j''}^2 }{ \mu_{j\phantom{''}}^2 }
    \bigg)^{(j''-j')(j'-j)}
    \Bigg(
    \bigg(
    \frac{ \mu_{j''}^2 }{ \mu_{j\phantom{''}}^2 }
    \bigg)^{j'-j}
    \bigg(
    \frac{ \mu_{j\phantom{'}}^2 }{ \mu_{j'}^2 }
    \bigg)^{j''-j}
    \Bigg)^{k + \frac{d-4}{2} }
\no \\[1.5ex]
    & \qquad \qquad \qquad
    \leq
    \frac{
    \begin{aligned}
    \Big(
    \mu_{j''}^{2(k+j''-j')+d-4} \,
    \big(
    b_{\mfn_j;k+j''-j',j''}
    &+ \Delta b_{\mfn_j;k+j''-j',j''}
    \big)
    \Big)^{j'-j}
    \\[-0.5ex]
    {} \times {}
    \Big(
    \mu_{j \vphantom{j'}}^{2(k+j-j')+d-4} \,
    \big(
    b_{\mfn_j;k+j-j',j}
    &+ \Delta b_{\mfn_j;k+j-j',j}
    \big)
    \Big)^{j''-j'}
    \end{aligned}
    }
    {
    \Big(
    \mu_{j'}^{2k+d-4} \,
    \big(
    b_{\mfn_j;k,j'}
    + \Delta b_{\mfn_j;k,j'}
    \big)
    \Big)^{j''-j}
    }
    \, .
\end{align}
The left-hand side is positive for all finite $k$. The $k \to \infty$ limit of the right-hand side can be computed using~\eqref{eq:klim-stu-3} and~\eqref{eq:klim-stu-6}. The result is finite and positive. Thus, the same limit of the left-hand side must also be finite. Hence, the quantity inside the large parentheses in~\eqref{eq:stu-SMC-ineq} must be less than or equal to one. Therefore,
\begin{align}
\label{eq:stu-SMC}
    \bigg(
    \frac{ \mu_{j''}^2 }{ \mu_{j\phantom{''}}^2 }
    \bigg)^{j'-j}
    \leq
    \bigg(
    \frac{ \mu_{j'}^2 }{ \mu_{j\phantom{'}}^2 }
    \bigg)^{j''-j}
    \, .
\end{align}
This completes our proof of the $stu$-symmetric SMC.

\subsection{Saturating the \texorpdfstring{$stu$}{stu}-SSC and \texorpdfstring{$stu$}{stu}-SMC}

We conclude this section by discussing the saturation of the $stu$-SSC and $stu$-SMC bounds.

\subsubsection{Saturating the \texorpdfstring{$stu$}{stu}-SSC}

The $stu$-SSC is a simple linear inequality, and it is trivially saturated by $J_n = 2\ceil{\mcJ_n/2}$ for all~${n \geq 1}$ with $m_n$ in the weakly-coupled spectrum. In this case, the maximum spins at each mass level increase linearly as $J_n = 2n+\hat{\jmath}-2$, where $\hat{\jmath} = \max(2\ceil{\alpha/2},J_0+2)$ is even, so that the masses obey ${m_n = \mu_{2n+\hat{\jmath}-2}}$. In summary, when the $stu$-SSC is saturated, the maximum spins increase by two at each subsequent mass level beginning with spin-$\hat{\jmath}$ at mass $m_1$.

The closed superstring spectrum in \autoref{fig:regge} saturates the $stu$-SSC with ${J_0 = \alpha = 2}$ and ${J_n = 2n+2}$. We revisit this observation in \autoref{sec:SYMSUGRA}.

\subsubsection{Saturating the \texorpdfstring{$stu$}{stu}-SMC}

The $stu$-SMC is more complicated because~\eqref{eq:stu-SMC} is highly non-linear. Moreover, the $stu$-SMC only constrains the set of masses $\mu_j$ at which a spin-$j$ state is first exchanged. In general, this set is a subset of the full set of masses $m_n$. That is,
\begin{align}
\label{eq:subset-stu}
    \{ \mu_j : \text{even } j \geq \hat{\jmath} \}
    \subset
    \{ m_n : n \geq 1 \}
    \, .
\end{align}
The smallest spin for which the $stu$-SMC applies is $\hat{\jmath} = \max(2\ceil{\alpha/2},J_0+2)$, and for all even $j \geq \hat{\jmath}$, the following masses saturate~\eqref{eq:stu-SMC}:
\begin{align}
\label{eq:stu-SMC-sol}
    \mu_j^2
    =
    \mu_{ \hat{\jmath} }^2 \,
    \bigg(
    \frac{ \mu_{ \hat{\jmath}+2 }^2 }
         { \mu_{ \hat{\jmath} \phantom{+2} }^2 }
    \bigg)^{ \frac{1}{2} (j-\hat{\jmath}) }
    \, .
\end{align}
Since we are free to measure masses in units of $\mu_{ \hat{\jmath} }$, this expression has one free parameter, the ratio of masses $\chi = \mu_{ \hat{\jmath}+2}^2 / \mu_{ \hat{\jmath}}^2$. As a consequence of the $stu$-SSC, these masses are ordered by $0 < \mu_{\hat{\jmath}} < \mu_{\hat{\jmath}+2}$, so we must have $\chi > 1$. Hence, $\chi \in (1, \infty)$ parametrizes a set of masses~$\mu_j$ which saturate~\eqref{eq:stu-SMC}.

If the masses $\mu_j$ saturate the $stu$-SMC, then we can also bound a subset of the couplings. If each $\mu_j$ is given by~\eqref{eq:stu-SMC-sol}, then the quantity inside the large parentheses in~\eqref{eq:stu-SMC-ineq} is now equal to one. The $k \to \infty$ limit of this expression can be computed using~\eqref{eq:klim-stu-3} and~\eqref{eq:klim-stu-6} and yields the following non-linear inequality involving both the masses and couplings:
\begin{align}
    \bigg(
    \frac{ \mu_{j''}^2 }{ \mu_{j\phantom{''}}^2 }
    \bigg)^{(j''-j')(j'-j)}
    \leq
    \frac{
    \big(
    g_{\mfn_{j''},j''}^{2} \, v_{j'',j''}^{(d)}
    \big)^{j'-j}
    \big(
    g_{\mfn_{j},j}^{2} \, v_{j,j}^{(d)}
    \big)^{j''-j'}
    }
    {
    \big(
    g_{\mfn_{j'},j'}^{2} \, v_{j',j'}^{(d)}
    \big)^{j''-j}
    }
    \, .
\end{align}
Choosing $j=\hat{\jmath}$ and $j'=\hat{\jmath}+2$, redefining ${j'' \mapsto j}$, and then using~\eqref{eq:stu-SMC-sol} to write the ratio of masses in terms of the parameter $\chi$, this inequality becomes
\begin{align}
\label{eq:stu-SMC-gineq}
    g_{\mfn_{j},j}^{2} \,
    v_{j,j}^{(d)}
    \geq
    g_{\mfn_{\hat{\jmath}},\hat{\jmath}}^{2} \,
    v_{\hat{\jmath},\hat{\jmath}}^{(d)} \,
    \chi^{\frac{1}{2}(j-\hat{\jmath})(j-\hat{\jmath}-2)} \,
    \xi^{\frac{1}{2}(j-\hat{\jmath})}
    \, ,
\end{align}
where we have defined
\begin{align}
    \xi
    =
    \frac{ g_{\mfn_{\hat{\jmath}+2},\hat{\jmath}+2}^{2} }
         { g_{\mfn_{\hat{\jmath}},\hat{\jmath}}^{2} \hfill }
    \frac{ v_{\hat{\jmath}+2,\hat{\jmath}+2}^{(d)} }
         { v_{\hat{\jmath},\hat{\jmath}}^{(d)} \hfill }
    \, .
\end{align}
The parameter $\xi > 0$ is strictly positive since the two couplings $g_{\mfn_{\hat{\jmath}},\hat{\jmath}}^{2}$ and $g_{\mfn_{\hat{\jmath}+2},\hat{\jmath}+2}^{2}$ are strictly positive as discussed in \autoref{sec:intro}. The final inequality~\eqref{eq:stu-SMC-gineq} is vacuously true for both ${j = \hat{\jmath}}$ and~${j = \hat{\jmath}+2}$, so it holds for all even ${j \geq \hat{\jmath}}$.

Now, we do not know of any theories whose mass spectrum saturates the $stu$-SMC, but we can prove that such a theory is inconsistent if particles of arbitrarily high spin appear in the weakly-coupled spectrum. If this theory were consistent, then its Wilson coefficients would be finite quantities. Let us consider the Wilson coefficient $a_{1;\hat{\jmath},0}^{(stu)}$ using the dispersion relation~\eqref{eq:disprel-stu}:
\begin{align}
    a_{1;\hat{\jmath},0}^{(stu)}
    &=
    \sum_{n \geq 1}
    \frac{2}{m_n^{2\hat{\jmath}+d-4}}
    \sum_{\substack{ j = 0 \\ \mathclap{\text{even}}\mathstrut}}^{ J_n }
    g_{n,j}^{2}
    +
    \frac{2}{\Lambda^{2\hat{\jmath}+d-4}}
    \int_0^1
    \mathrm{d}z \,
    z^{\hat{\jmath}-1+\frac{d-4}{2}}
    \sum_{\substack{ j = 0 \\ \mathclap{\text{even}}\mathstrut}}^{ \infty }
    \rho_j \big( \Lambda^2 z^{-1} \big)
    \, .
\end{align}
Here we have used $w_{j;\hat{\jmath},0}^{(d)}=2$. Let us derive a lower bound on this expression. Since $\rho_j(s) \geq 0$ and $g_{n,j}^2 \geq 0$, we can drop the integral and then use~\eqref{eq:subset-stu} to restrict the sum over masses from the set of $m_n$ to the subset of $\mu_j$ to find
\begin{align}
    a_{1;\hat{\jmath},0}^{(stu)}
    & \geq
    \sum_{\substack{ j \geq \hat{\jmath} \\ \mathclap{\text{even}}\mathstrut}}
    \frac{2}{\mu_j^{2\hat{\jmath}+d-4}} \,
    g_{\mfn_j,j}^{2}
    \, .
\end{align}
Using~\eqref{eq:stu-SMC-sol} to write each $\mu_j$ in terms of the parameter $\chi > 1$ and \eqref{eq:stu-SMC-gineq} to bound the couplings, we find
\begin{align}
    a_{1;\hat{\jmath},0}^{(stu)}
    \geq
    2 \,
    \mu_{\hat{\jmath}}^{-2\hat{\jmath}+4-d} \,
    g_{\mfn_{\hat{\jmath}},\hat{\jmath}}^{2} \,
    v_{\hat{\jmath},\hat{\jmath}}^{(d)}
    \sum_{\substack{ j \geq \hat{\jmath} \\ \mathclap{\text{even}}\mathstrut}}
    \chi^{\frac{1}{2}(j-\hat{\jmath})(j-3\hat{\jmath}+2-\frac{d}{2})} \,
    \xi^{\frac{1}{2}(j-\hat{\jmath})} \,
    \frac{ 1 }{ v_{j,j}^{(d)} }
    \, .
\end{align}
If the sum over $j$ is infinite (that is, if the theory contains particles of arbitrarily high spin in its weakly-coupled spectrum), then this sum diverges by the ratio test since $\chi > 1$ and $\xi > 0$. In this case, $a_{1;\hat{\jmath},0}^{(stu)}$ diverges, and the theory is inconsistent. Therefore, the $stu$-SMC cannot be saturated for an infinite set of spins, which implies that the $stu$-SMC~\eqref{eq:stu-SMC} is not an optimal bound. 

\section{Super-gluon and super-graviton amplitudes}
\label{sec:SYMSUGRA}

In this section, we briefly discuss how the Sequential Spin Constraints (SSC) and Sequential Mass Constraints (SMC) might be generalized to the four-point amplitudes describing the scattering of any states in the massless supermultiplets of super Yang-Mills (SYM) theories or supergravity (SUGRA) theories. To this end, we consider any EFT whose massless sector is described by maximally supersymmetric SYM or SUGRA. We dub these theories SYM-EFT and SUGRA-EFT, respectively. This particular bootstrap problem was recently studied in~\cite{Berman:2024wyt, Berman:2023jys, Haring:2023zwu, Albert:2024yap}. The ten-dimensional Type~I and~II superstring theories provide canonical UV-complete solutions, but our conclusions hold for any putative maximal SYM-EFT or SUGRA-EFT that may exist, UV-complete or not.

\subsection{Super-gluons}

Let us first consider a SYM-EFT. The massless sector is given by the super-gluon multiplet. The color-stripped tree-level field theory amplitude which describes the scattering of any four massless particles in this supermultiplet can be written as
\begin{align}
    A_\text{SYM-tree}
    &=
    g_\text{YM}^2 \,
    \mcF^4 \,
    \frac{1}{su}
    \, ,
\end{align}
where $\mcF^4$ is a kinematic prefactor which is determined by maximal supersymmetry and which contains the information about the particular states being scattered~\cite{Arkani-Hamed:2022gsa}. This kinematic prefactor is a function of the momenta and polarizations of the external states.

Using supersymmetric Ward identities~\cite{Albert:2024yap, Elvang:2013cua}, one can show that the same amplitude in a generic SYM-EFT depends on a single $su$-symmetric scalar function $f(s,u)$,
\begin{align}
\label{eq:ASYMEFT}
    A_\text{SYM-EFT}
    &=
    g_\text{YM}^2 \,
    \mcF^4 \,
    f(s,u)
    \, ,
\end{align}
whose low-energy expansion is given by
\begin{align}
    f(s,u)
    &=
    \frac{1}{su}
    +
    \text{analytic terms}
    \, .
\end{align}
This scalar function has a positive partial wave expansion like~\eqref{eq:partial}~\cite{Albert:2024yap, Berman:2023jys, Berman:2024wyt, Haring:2023zwu, Arkani-Hamed:2022gsa} and determines all the four-point scattering amplitudes for the super-gluon multiplet. All information about the helicities of the external states is encoded in $\mcF^4$.

Tree-level open superstring theory provides a canonical example of~\eqref{eq:ASYMEFT}, the Veneziano amplitude~\cite{Veneziano:1968yb} (with vanishing Regge intercept and Regge slope $\alpha'$):
\begin{align}
    f_\text{open}(s,u)
    &=
    (\alpha')^2 \,
    \frac{ \Gamma(-\alpha's) \, \Gamma(-\alpha'u) }
         { \Gamma(1-\alpha's-\alpha'u) }
    \, .
\end{align}
In the generic case, we can multiply the scalar function $f(s,u)$ by ${t=-(s+u)}$ to produce a stripped-amplitude $A^{(su)}(s,u)$:
\begin{align}
\label{eq:subamp-su}
    A^{(su)}(s,u)
    =
    -(s+u) \, f(s,u)
    =
    - \bigg( \frac{1}{s} + \frac{1}{u} \bigg)
    + \text{analytic terms}
    \, .
\end{align}
In~\cite{Albert:2024yap}, it was argued that this stripped-amplitude should obey the usual $su$-symmetric bootstrap assumptions described in~\autoref{sec:intro}. If we further assume that $A^{(su)}(s,u)$ exchanges a finite number of spins at each weakly-coupled mass level, then we can apply the $su$-SSC and $su$-SMC to produce bounds on the massive spectrum in any effective theory of super-gluons!

The consequences of the $su$-SSC require some care. Because the kinematic prefactor in~\eqref{eq:ASYMEFT} scales with $u$ like $\mcF^4 \sim u^2$ (for at least some choice of external states), the full SYM-EFT amplitude will be related to the stripped-amplitude by $A_\text{SYM-EFT} \sim u^1 \cdot A^{(su)}(s,u)$. This extra factor of $u$ then modifies the spin spectrum. At each mass level, the full SYM-EFT amplitude $A_\text{SYM-EFT}$ will exchange spins one unit higher than those of the stripped-amplitude $A^{(su)}(s,u)$.

Let us now determine the appropriate values of $J_0$ and $\alpha$ for the stripped-amplitude. From~\eqref{eq:subamp-su}, we see that the stripped-amplitude describes the exchange of massless scalars so that ${J_0 = 0}$. To determine $\alpha$, we look to the example of the open superstring. In this case, ${-(s+u) \, f_\text{open}(s,u) \sim s^{\alpha' u}}$ (with fixed ${u < 0}$) is polynomially-bounded by~$s^{0}$ so that ${\alpha=0}$. Thus, we assume that ${\alpha = 0}$ in the generic case as well. This UV behavior is marginally better than that of tree-level field theory since ${-(s+u)\frac{1}{su} \sim s^0}$ is only bounded by~$s^{1}$.

With ${J_0 = \alpha = 0}$, the $su$-SSC implies that the spins of the massive states exchanged in the stripped-amplitude are bounded by ${J_n \leq \max(0,J_1,J_2,\dots,J_{n-1})+1}$. Shifting by one, we conclude that the spins exchanged in the full SYM-EFT amplitude are bounded~by
\begin{align}
    J_n
    \leq
    \max(1,J_1,J_2,\dots,J_{n-1})+1
    \, .
\end{align}
This spin bound holds for the states exchanged in any four-gluon SYM-EFT amplitude. The open superstring spectrum saturates this bound with ${J_n = n+1}$ for all $n \geq 0$.

The $su$-SMC simply implies that the masses of exchanged states in any SYM-EFT must satisfy~\eqref{eq:su-SMC}. In particular, the open superstring spectrum is given by ${\alpha' m_n^2 = n}$, which satisfies (but does not saturate) the SMC because
\begin{align}
\label{eq:n0n1n2}
    \bigg( \frac{n_2}{n_0} \bigg)^{n_1-n_0}
    <
    \bigg( \frac{n_1}{n_0} \bigg)^{n_2-n_0}
\end{align}
for all integers $n_2 > n_1 > n_0 \geq 1$.

\subsection{Super-gravitons}

Now let us consider a SUGRA-EFT. The massless sector is given by the super-graviton multiplet. The tree-level field theory amplitude which describes the scattering of any four massless particles in this supermultiplet can be written as
\begin{align}
    A_\text{SUGRA-tree}
    &=
    \kappa^2 \,
    \mcR^4 \,
    \frac{1}{stu}
    \, ,
\end{align}
where $\mcR^4$ is a kinematic prefactor (like $\mcF^4$) which is determined by maximal supersymmetry and which contains the information about the particular states being scattered~\cite{Arkani-Hamed:2022gsa,Haring:2023zwu}.

Using supersymmetric Ward identities~\cite{Albert:2024yap, Elvang:2013cua}, one can show that the same amplitude in a generic SUGRA-EFT depends on a single $stu$-symmetric scalar function $g(s,t,u)$,
\begin{align}
\label{eq:ASUGRAEFT}
    A_\text{SUGRA-EFT}
    &=
    \kappa^2 \,
    \mcR^4 \,
    g(s,t,u)
    \, .
\end{align}
This scalar function is related to dilaton scattering, has a positive partial wave expansion like~\eqref{eq:partial}~\cite{Albert:2024yap, Berman:2023jys, Berman:2024wyt, Haring:2023zwu, Arkani-Hamed:2022gsa}, and determines all the four-point scattering amplitudes for the super-graviton multiplet.

Tree-level closed superstring theory provides a canonical example of~\eqref{eq:ASUGRAEFT}, the Virasoro-Shapiro amplitude~\cite{Virasoro:1969me, Shapiro:1970gy} (with vanishing Regge intercept and Regge slope $\frac{\alpha'}{4}$):
\begin{align}
    g_\text{closed}(s,t,u)
    &=
    \Big(
    \frac{\alpha'}{4}
    \Big)^3
    \frac{ \Gamma(-\frac{\alpha'}{4} s) \,
           \Gamma(-\frac{\alpha'}{4} t) \,
           \Gamma(-\frac{\alpha'}{4} u) }
         { \Gamma(1+\frac{\alpha'}{4} s) \,
           \Gamma(1+\frac{\alpha'}{4} t) \,
           \Gamma(1+\frac{\alpha'}{4} u) }
    \, .
\end{align}
In the generic case, we can multiply the scalar function $g(s,t,u)$ by $\frac{1}{4} (s^2+t^2+u^2)^2$ to produce a stripped-amplitude $A^{(stu)}(s,u)$:
\begin{align}
\label{eq:subamp-stu}
    A^{(stu)}(s,u)
    =
    \frac{1}{4} (s^2+t^2+u^2)^2 \, g(s,t,u)
    =
    \bigg( \frac{tu}{s} + \frac{us}{t} + \frac{ut}{s} \bigg)
    + \text{analytic terms}
    \, .
\end{align}
In~\cite{Albert:2024yap}, it was argued that this stripped-amplitude should obey the usual $stu$-symmetric bootstrap assumptions described in~\autoref{sec:intro}. In fact, this stripped-amplitude is precisely the four-dilaton amplitude stripped of the gravitational coupling $\kappa^2$. If we further assume that $A^{(stu)}(s,u)$ exchanges a finite number of spins at each weakly-coupled mass level, then we can apply the $stu$-SSC and $stu$-SMC to produce bounds on the massive spectrum in any effective theory of super-gravitons!


Let us now determine the appropriate values of $J_0$ and $\alpha$ for the stripped-amplitude. From~\eqref{eq:subamp-stu}, we see that the stripped-amplitude describes the exchange of massless gravitons so that ${J_0 = 2}$. To determine $\alpha$, we look to the example of the closed superstring. In this case, ${\frac{1}{4}(s^2+t^2+u^2)^2 \, g_\text{closed}(s,t,u) \sim s^{\frac{1}{2}\alpha' u+2}}$ (with fixed ${u < 0}$) is polynomially-bounded by~$s^{2}$ so that ${\alpha=2}$. Thus, we assume that $\alpha = 2$ in the generic case as well. This UV behavior is marginally better than that of tree-level field theory since ${\frac{1}{4}(s^2+t^2+u^2)^2 \frac{1}{stu} \sim s^{2}}$ is only bounded by~$s^{3}$.

With ${J_0 = \alpha = 2}$, the $stu$-SSC implies that the spins of the massive states exchanged in both the stripped-amplitude and the full SUGRA-EFT amplitude are bounded by 
\begin{align}
    J_n
    \leq
    \max(2,J_1,J_2,\dots,J_{n-1})+2
    \, .
\end{align}
This spin bound holds for the states exchanged in any four-graviton SUGRA-EFT amplitude. The closed superstring spectrum saturates this bound with ${J_n = 2n+2}$ for all $n \geq 0$.

The $stu$-SMC simply implies that the masses of exchanged states in any SUGRA-EFT must satisfy~\eqref{eq:stu-SMC}. In particular, the closed superstring spectrum is given by ${\frac{1}{4} \alpha' m_n^2 = n}$, which satisfies (but does not saturate) the SMC because of~\eqref{eq:n0n1n2}.

\section{Discussion}
\label{sec:disc}

In this paper, we proved the Sequential Spin Constraints (SSC) and Sequential Mass Constraints (SMC) for generic $su$- and $stu$-symmetric four-point massless scalar amplitudes (with or without gravity) using a standard set of S-matrix bootstrap assumptions. We discussed our assumptions and the detailed form of our results in \autoref{sec:intro}. We then proved the SSC and SMC in the following two sections. Our proof of the $su$-symmetric case simplifies that of~\cite{Berman:2024kdh}, and our proof of the $stu$-symmetric case is a new result. Finally, in \autoref{sec:SYMSUGRA} we discussed the generalization of our results to maximally supersymmetric super-gluon and super-graviton amplitudes.

It is important to note that our analysis of the $su$-symmetric amplitudes only made use of fixed-$u$ dispersion relations. One can also derive an independent set of fixed-$t$ dispersion relations for these amplitudes. (The fixed-$u$ and fixed-$t$ dispersion relations are identical in the $stu$-symmetric case.) While both sets of dispersion relations provide independent constraints at finite $k$~\cite{Berman:2023jys, Berman:2024wyt}, our proof ultimately relies on the $k \to \infty$ limit. If we were to proceed through \autoref{sec:su} with the fixed-$t$ dispersion relations, we would end up with the same mass and spin bounds as we found with the fixed-$u$ approach. In fact, the fixed-$t$ dispersion relations for $su$-symmetric amplitudes resemble the $stu$-symmetric dispersion relations from which we derive identical mass and spin bounds.

While our precise statement of the SSC is new to the literature, a result quite similar to the $su$-SSC appeared in~\cite{Caron-Huot:2016icg} at the advent of the modern S-matrix bootstrap program. Specifically, the authors of~\cite{Caron-Huot:2016icg} proved an asymptotic version of the SSC for meromorphic $su$-symmetric amplitudes, but they did not prove the statement that the lightest spin\nobreakdash-$j$ state must be lighter than the lightest spin-$(j+1)$ state for all $j$. Therefore, our statement of the $su$-SSC provides an infinite number of analytic constraints which are not technically implied by the asymptotic result of~\cite{Caron-Huot:2016icg}: the lightest state with any given spin must always appear before states with higher spins. Additionally, the argument of~\cite{Caron-Huot:2016icg} makes use of an analytic continuation of its meromorphic amplitudes into the unphysical regime with large and positive ${s,u \gg 0}$, highlighting the interplay between the analytic properties of an amplitude and its spectrum. In this paper, we only assume meromorphicity below the cut-off ${|s|, |u| < \Lambda^2}$ to allow for loops of massive particles and non-meromorphic tree-level amplitudes with accumulation point spectra such as the Coon amplitude~\cite{Coon:1969yw, Figueroa:2022onw} (which appeared as a counterexample in~\cite{Caron-Huot:2016icg}) and its hypergeometric cousins~\cite{Cheung:2023adk}.

To our knowledge, the SMC was first described in the prelude to this work~\cite{Berman:2024kdh}, and no similar statements about the masses of spinning states appear in the bootstrap literature. It is, however, unsurprising that such constraints follow from the bootstrap assumptions. Polynomial boundedness alone places strict constraints on the spectrum. Roughly, the exchange of each massive spin-$j$ particle contributes to the amplitude as
\begin{align}
    \mcP^{(d)}_j \Big( 1+\frac{2s}{m^2} \Big)
    \sim s^j
\end{align}
in the Regge limit. For the amplitude to be polynomially-bounded by $s^\alpha$, the contributions with~${j \geq \alpha}$ need to resum. The SMC provides a necessary bound on the masses of these higher-spin states so that their contributions to the amplitude can properly resum. This resummation is a defining property of string theory amplitudes~\cite{Cappelli:2012cto} and must be similarly exhibited by all healthy UV completions.

We hope the SSC and SMC can serve as a guide to help probe the space of physically consistent theories. Indeed, all the analytic expressions for four-point amplitudes which satisfy our bootstrap assumptions (both from the old S-matrix bootstrap days and from the modern program) obey the SSC and SMC~\cite{Veneziano:1968yb, Virasoro:1969me, Lovelace:1968kjy, Figueroa:2022onw, Shapiro:1969km, Shapiro:1970gy, Coon:1969yw,Huang:2022mdb, Geiser:2022exp, Geiser:2022icl, Cheung:2022mkw, Cheung:2023adk, Cheung:2023uwn, Cheung:2024uhn, Cheung:2024obl}. In fact, the SSC implicitly appears within the infinite product constructions of~\cite{Cheung:2022mkw, Geiser:2022exp}.

There are, however, a class of amplitudes which have appeared in the modern S-matrix bootstrap program and which do not satisfy the SSC and SMC: the infinite spin towers.

\subsection{Infinite spin towers}

The infinite spin tower (IST) amplitudes, first introduced in~\cite{Caron-Huot:2020cmc}, satisfy all of the bootstrap assumptions discussed in \autoref{sec:intro} except for finite spin exchange. Hence, the ISTs evade the conclusions of this paper. The ISTs come in $su$- and $stu$-symmetric varieties:
\begin{align}
\label{eq:IST}
    A^{(su\text{-IST})}(s,u)
    &=
    g^2 \frac{m^4}{(s-m^2)(u-m^2)}
    \, ,
\no \\
    A^{(stu\text{-IST})}(s,u)
    &=
    g^2 \frac{m^6}{(s-m^2)(t-m^2)(u-m^2)}
    \, .
\end{align}
With our conventions, the coupling has mass dimension $[g^2] = 4-d$. The $su$-IST is polynomially-bounded by $s^{0}$, and the $stu$-IST is polynomially-bounded by $s^{-1}$.

The residues of these amplitudes at ${s=m^2}$ are non-polynomial in $u$ but can be expanded into an infinite sum of Gegenbauer polynomials with positive coefficients. Thus, the spectrum of each IST consists of an infinite set of particles with mass $m$ and all possible spins (${j=0,1,2,\dots}$ for the $su$-IST and ${j=0,2,4,\dots}$ for the $stu$-IST). This spectrum, with an infinite set of spins at a single mass, manifestly violates the SSC. The SMC, however, is still true for the IST spectrum since ${1^{j'-j} \leq 1^{j''-j}}$ for all ${j'' \geq j' \geq j}$. In fact, the ISTs can be viewed as a marginal solution of the SSC and SMC with a degenerate vertical Regge trajectory and all spins at a single mass.

While the ISTs violate the assumption of finite spin exchange, they otherwise appear to be healthy four-point amplitudes. In the weak-coupling limit, they are unitary (positive), polynomially-bounded, and analytic in $s$ away from their obvious simple poles. However, it is not yet known if these tree-level four-point amplitudes can be generalized to higher points and higher loops. Moreover, the authors of~\cite{Acanfora:2023axz} claim in their Appendix~C that the tree-level $stu$-IST violates non-linear unitarity at finite coupling.

In any case, the tree-level four-point ISTs regularly appear at the boundaries of allowed theory space, and they are notoriously difficult to exclude from numerical bootstrap studies~\mbox{\cite{Caron-Huot:2020cmc, Caron-Huot:2021rmr, Albert:2022oes, Albert:2024yap, Berman:2024wyt, Berman:2024eid}}. The precise role of ISTs in the S\nobreakdash-matrix bootstrap program is still an open question.

As we suggested in our discussion of finite spin exchanged in \autoref{sec:intro}, it may be reasonable to conclude that the ISTs correspond to a non-local microscopic theory. To this end, it was recently proposed in~\cite{Wang:2024jhc} that the $stu$-IST describes tree-level chiral string theory, a variation of worldsheet string theory in which the transformation $\alpha' \mapsto -\alpha'$ is applied to the right-handed string modes. If this proposal holds true, then we suspect that one could trace the $stu$-IST's violation of the SSC back to some unphysical property of the chiral worldsheet. Furthermore, if the ISTs are ultimately inconsistent (as evidenced by their non-unitarity at finite coupling~\cite{Acanfora:2023axz}), then we suspect that this inconsistency may manifest as some loop-level anomaly in chiral string theory.

\subsection{Future directions}

Let us conclude with some future directions of research.

\paragraph{Bounding couplings}

While we have derived bounds on the masses and spins of particles, we have said nothing of their couplings (other than the assumption of weak-coupling). Only when the mass spectrum saturated the SMC could we write any non-trivial inequalities involving couplings, namely~\eqref{eq:su-SMC-gineq} and~\eqref{eq:stu-SMC-gineq}. In this case, we were able to rule out the existence of amplitudes with an infinite set of weakly-coupled states whose masses saturate the SMC. More general coupling bounds could perhaps be used to rule out other spectra.

In any case, couplings are of great interest both to experimental particle physicists and to the bootstrap community. Recent bootstrap studies of QCD have derived rigorous numerical bounds on couplings (normalized by a particular Wilson coefficient)~\cite{Albert:2022oes} and correlations between various couplings~\cite{Albert:2023seb}. These numerical results are beyond the scope of this paper, but analytic techniques may still yield interesting results. For instance, more general inequalities than those studied in this paper can be derived using Hankel matrix techniques~\cite{Bellazzini:2020cot, Arkani-Hamed:2020blm, Huang:2022mdb, Berman:2024kdh, Wan:2024eto}. Perhaps these inequalities place constraints on the couplings.

\paragraph{Regge trajectories}

It was recently shown in~\cite{Eckner:2024pqt}, that $su$-symmetric amplitudes with a finite number of Regge trajectories are inconsistent and that consistent amplitudes must have an infinite number of daughter trajectories. String amplitudes have an infinite number of Regge trajectories, and the spectrum of real-world QCD exhibits daughter trajectories in accordance with~\cite{Eckner:2024pqt}. However, single-Regge-trajectory amplitudes regularly appear at the boundaries of theory space in the numerical S\nobreakdash-matrix bootstrap~\mbox{\cite{Caron-Huot:2020cmc, Caron-Huot:2021rmr, Albert:2022oes, Albert:2024yap, Berman:2024wyt, Berman:2024eid}}. One example is the infinite spin tower, which has a single degenerate Regge trajectory with infinite slope (i.e.\ all spins at a single mass).

Unfortunately, there is no hint of the result of~\cite{Eckner:2024pqt} in this paper. While we have excluded the IST by assumption, the SSC and SMC themselves do not prohibit the existence of a finite number of Regge trajectories. In the conventions of this paper, the existence of an infinite number of Regge trajectories means that there must be an infinite number of non-zero couplings ${g_{n,j}^2 > 0}$ at each $j$. Deeper insights into the couplings may thus shed light on the Regge trajectories. In future work, we hope to explicitly study the properties of Regge trajectories along with the couplings. An immediate goal is the generalization of~\cite{Eckner:2024pqt} to $stu$-symmetric amplitudes.

\paragraph{Beyond weak-coupling}

Throughout this work, we have only worked with weakly-coupled theories. Correspondingly, we have only used the positivity relation~\eqref{eq:positivity} rather than the full non-linear unitarity relation~\eqref{eq:unitarity}. Full unitarity is a much stronger constraint than positivity, but one must go beyond tree-level to go beyond positivity. Some ideas for incorporating loops into the bootstrap are presented in~\cite{Arkani-Hamed:2020blm, Haring:2023zwu, Bellazzini:2021oaj}. Systematically including loop-level effects would allow us to harness the full power of unitarity. A finite-coupling bootstrap could also kill the infinite spin tower (without assuming finite spin exchange) as discussed above.

\paragraph{Beyond identical massless scalars}

We also hope to extend our results to other four-point amplitudes. We suspect that similar bounds can be derived for the weakly-coupled states exchanged between four-point photon, gluon, graviton, or multi-scalar amplitudes. We also believe that a generalization to massive scalar amplitudes may be in reach. While the set-up of this paper allows for these states to be exchanged as virtual particles between identical massless scalars, it will be fruitful to consider them as external states. Each new set of bounds will further constrain the spectra of consistent effective theories, and we anticipate that any new bounds could be applied to predict the QCD spectrum as in~\cite{Berman:2024kdh}.

\paragraph{The Conformal Bootstrap}

Given that the essential features of our proof are crossing symmetry and the existence of positive moments, it would be interesting to understand if the same methods could be applied to the conformal bootstrap problem~\cite{Poland:2022qrs}. Similar bounds have already been derived in the conformal bootstrap literature using a variety of other techniques, e.g.~\cite{Fitzpatrick:2012yx, Komargodski:2012ek, vanRees:2024xkb}. Thus, it is possible that one could find analytic bounds on the conformal dimensions and spins of operators in consistent conformal field theories, analogous to the SSC and SMC. 

\subsection*{Acknowledgments}

We thank (in alphabetical order) Jan Albert, Paolo Di Vecchia, Henriette Elvang, Yue-Zhou Li, Loki Lin, Slava Rychkov, and Sasha Zhiboedov for many helpful comments and conversations. We also thank the Niels Bohr International Academy at the Niels Bohr Institute at the University of Copenhagen for hospitality during the final stages of this work. JB is supported in part by the US Department of Energy grant DE-SC0007859 and by a Cottrell SEED Award number CS-SEED-2023-004 from the Research Corporation for Science Advancement. NG is supported by a Leinweber Postdoctoral Research Fellowship and a Van Loo Postdoctoral Research Fellowship from the University of Michigan along with partial support from the US Department of Energy grant DE-SC0007859.

\bibliography{bib}

\end{document}